\documentclass[a4paper,fleqn,usenatbib]{mnras}

\usepackage{newtxtext,newtxmath}

\usepackage[T1]{fontenc}
\usepackage{ae,aecompl}

\usepackage{graphicx}	
\usepackage{amsmath}	
\usepackage{amssymb}	

\usepackage{relsize}

\usepackage{cuted}
\usepackage{cases}

\title[SILCC-Zoom: H$_2$ and CO-dark gas]{
SILCC-Zoom: H$_2$ and CO-dark gas in molecular clouds -- The impact of feedback and magnetic fields
}
  \author[D. Seifried et al.]
  {D.~Seifried,$^1$\thanks{seifried@ph1.uni-koeln.de} S.~Haid,$^1$ S.~Walch,$^1$ E.~M.~A.~Borchert,$^1$ T.~G.~Bisbas$^{1,2,3}$  \\
  $^1$Universit\"at zu K\"oln,I. Physikalisches Institut,  Z\"ulpicher Str. 77, 50937 K\"oln, Germany\\
  $^2$Department of Physics, Aristotle University of Thessaloniki, 54124 Thessaloniki, Greece\\
  $^3$National Observatory of Athens, Institute for Astronomy, Astrophysics, Space Applications and Remote Sensing, Penteli, 15236, Athens, Greece
  }

\date{Released 2019}

\pubyear{2019}

\begin{document}
\label{firstpage}
\pagerange{\pageref{firstpage}--\pageref{lastpage}}
\maketitle

\begin{abstract}
We analyse the CO-dark molecular gas content of simulated molecular clouds from the SILCC-Zoom project. The simulations reach a resolution of 0.1 pc and include H$_2$ and CO formation, radiative stellar feedback and magnetic fields. CO-dark gas is found in regions with local visual extinctions $A_\rmn{V, 3D} \sim$ 0.2 -- 1.5, number densities of 10 -- \mbox{10$^{3}$ cm$^{-3}$} and gas temperatures of few \mbox{10 K} -- \mbox{100 K}. CO-bright gas is found at number densities above \mbox{300 cm$^{-3}$} and temperatures below 50~K. The CO-dark gas fractions range from 40\% to 95\% and scale inversely with the amount of well-shielded gas ($A_\rmn{V, 3D}$ $\gtrsim$ 1.5), which is smaller in magnetised molecular clouds. We show that the density, chemical abundances and $A_\rmn{V, 3D}$ along a given line-of-sight cannot be properly determined from projected quantities. As an example, pixels with a projected visual extinction of $A_\rmn{V, 2D} \simeq$ 2.5 -- 5 can be both, CO-bright or CO-dark, which can be attributed to the presence or absence of strong density enhancements along the line-of-sight. By producing synthetic CO(1-0) emission maps of the simulations with RADMC-3D, we show that about 15 -- 65\% of the H$_2$ is in regions with intensities below the detection limit. Our clouds have $X_\rmn{CO}$-factors around \mbox{1.5 $\times$ 10$^{20}$ cm$^{-2}$ (K km s$^{-1}$)$^{-1}$} with a spread of up to a factor $\sim$ 4, implying a similar uncertainty in the derived total H$_2$ masses and even worse for individual pixels. Based on our results, we suggest a new approach to determine the H$_2$ mass, which relies on the availability of CO(1-0) emission and $A_\rmn{V, 2D}$ maps. It reduces the uncertainty of the clouds' overall H$_2$ mass to a factor of $\lesssim$ 1.8 and for individual pixels, i.e. on sub-pc scales, to a factor of $\lesssim$~3.
\end{abstract}





\section{Introduction}

Molecular clouds (MCs) are the densest structures of the interstellar medium (ISM) and are defined as those regions in which hydrogen exists in its molecular form, H$_2$. Due to the low temperatures of a few 10 K and its vanishing permanent dipole moment, H$_2$ and thus MCs are observable only indirectly e.g. by molecules which trace the presence of H$_2$. One of the most frequently used molecules is CO \citep[e.g.][and many more, but see \citealt{Dobbs14} for a review]{Wilson70,Scoville75,Larson81,Solomon87,Dame01,Bolatto13}.

However, it has been shown that CO requires a more efficient shielding of the interstellar radiation field (ISRF) than H$_2$ to form \citep{Dishoeck88,Wolfire10}. Thus, CO is not a perfect tracer of H$_2$ gas as it misses out a considerable fraction of molecular gas. In this context, the terminology of ``CO-poor'' or ``CO-dark'' gas, i.e. gas, where H$_2$ is present but no CO, was established \citep{Lada88,Dishoeck92,Grenier05}. Using gamma ray emission \citet{Grenier05} conclude that more than 30\% of the H$_2$ gas is CO-dark \citep[see also e.g.][]{Ackermann12,Donate17}.

Since then, CO-dark gas has been subject to a number of studies. By means of dust extinction measurements, CO-dark gas fractions of up to several 10\% were found \citep[e.g.][]{Lee12,Planck15}, similar to the results obtained from gamma ray emission studies. Furthermore, observations of the [C II] 158 $\mu$m line suggest the presence of CO-dark H$_2$ gas in order to consistently explain the observed line intensity \citep{Langer10,Langer14,Pineda13}. These findings were recently supported theoretically by \citet{Franeck18}, who show that up to 20\% of the [C II] line emission stems from the molecular phase. Moreover, emission of atomic carbon has been suggested to be a good tracer of CO-dark molecular gas \citep{Gerin00,Papadopoulos04,Offner14,Glover15,Li18,Clark19}.

More recently, other tracers have also been used to study CO-dark gas, in particular the hydroxyl radical OH \citep[e.g.][]{Crutcher93,Barriault10,Cotten12,Allen15,Li15,Li18b,Ebisawa19}. In a sub-pc resolution observation of OH, \citet{Xu16} find CO-dark gas fractions varying from 80\% to 20\% across the boundary of the Taurus molecular cloud. In addition, other molecules like HF, HCl and ArH$^+$ have been suggested to be able to probe the atomic and molecular hydrogen content of MCs and thus the amount of CO-dark gas \citep{Schilke95,Schilke14,Sonnentrucker10,Neufeld97,Neufeld05,Neufeld16}.

Taken together, these observations draw a clear picture of MCs where a significant fraction of H$_2$ is not detectable in CO. In order to still be able to infer the amount of H$_2$ gas from CO observations, a conversion factor from the observed CO luminosity into an H$_2$ column density has been established, the so-called ``$X_\rmn{CO}$-factor''. Its canonical value in the MilkyWay is assumed to be about \mbox{2 $\times$ 10$^{20}$ cm$^{-2}$ (K km s$^{-1}$)$^{-1}$} \citep[see e.g. the review by][]{Bolatto13}. However, there are significant cloud-to-cloud variations of $X_\rmn{CO}$ reported in the literature in both observations of galactic and extra-galactic MCs \citep[e.g.][]{Blitz80,Scoville87,Dame93,Strong96,Melchior00,Lombardi06,Nieten06,Leroy11,Smith12,Ripple13}. In addition, also metallicity variations affect the value of X$_\rmn{CO}$ \citep{Glover11,Shetty11a,Bolatto13}. All these variations imply uncertainties of a factor of a few in the masses of H$_2$ inferred from CO observations and thus, the $X_\rmn{CO}$-factor might be applicable only for an ensemble of clouds rather than individual clouds \citep{Kennicutt12}.

The large spread of the amount of CO-dark gas and the resulting $X_\rmn{CO}$-factor was confirmed by a number of recent numerical simulations of MC formation \citep{Glover11,Smith14,Duarte15,Glover16,Richings16a,Richings16b,Szucs16,Gong18,Li18}. However, there are two stringent constraints on the accuracy of such numerical approaches: First, due to the highly turbulent structure of MCs and the associated mixing of molecules \citep{Glover10,Valdivia16,Seifried17}, the chemical evolution of such clouds has to be modelled on-the-fly in the simulations in order to obtain an accurate picture of their -- partly non-equilibrium -- chemical state. Second, it can be shown numerically \citep{Seifried17} \textit{and} analytically \citep{Joshi19} that a very high spatial resolution of $\lesssim$ 0.1 pc is required to obtain accurate and converged chemical abundances. In case a numerical simulation does not reach this resolution or the chemical (non-equilibrium) abundances are not modelled on-the-fly, inferred fractions of CO-dark gas and values of the $X_\rmn{CO}$-factor have to be considered with caution. In addition, a potential complication arises from highly idealized initial conditions, which do not match the full complexity of real MCs \citep{Rey15}.

So far, only a few simulations match the aforementioned requirements. Moreover, the impact of stellar radiative feedback and of magnetic fields on the amount of CO-dark gas and the properties of CO emission has obtained very little attention so far. In \citet{Seifried17,Seifried19} and \citet{Haid19} we present some of the first numerical simulations of MC formation which include an on-the-fly chemical network for H$_2$ and CO, high spatial resolution ($\sim$ 0.1 pc), a larger-scale, galactic environment for realistic initial conditions, magnetic fields and stellar feedback. In the following we will use these simulations to investigate the impact of feedback and magnetic fields on CO-dark gas and the observable CO emission in detail.

The structure of the paper is as follows: First, we describe the initial conditions and numerical methods used for the MC simulations (Section~\ref{sec:numerics}). We then present our results and discuss the amount and distribution of CO-dark gas (Sections~\ref{sec:dgf_total} and~\ref{sec:distribution}). Next, we investigate projection effects (Section~\ref{sec:2dmaps}) and the effect of CO-dark gas on the CO emission and the $X_\rmn{CO}$-factor (Section~\ref{sec:CO}). Finally, in Section~\ref{sec:newapproach}, we develop a new approach to determine the H$_2$ mass in MCs with a higher accuracy than via the $X_\rmn{CO}$-factor before we conclude in Section~\ref{sec:conclusion}.

\section{Numerics and initial conditions}
\label{sec:numerics}

We present results of the SILCC-Zoom simulations of MC formation \citep{Seifried17}. The simulations are preformed within the SILCC project \citep[see][for details]{Walch15,Girichidis16} and make use of the zoom-in technique discussed in \citet{Seifried17}. The simulations are performed with the adaptive mesh refinement code FLASH 4.3 \citep{Fryxell00,Dubey08} and use a magneto-hydrodynamics (MHD) solver which guarantees positive entropy and density \citep{Bouchut07,Waagan09}. We model the chemical evolution of the interstellar medium (ISM) using a simplified chemical network for H$^+$, H, H$_2$, C$^+$, CO, e$^-$, and O \citep{Nelson97,Glover07,Glover10}, which also follows the thermal evolution of the gas including the most important heating and cooling processes. We do not apply any particular treatment for compressive shocks like shattering or sputtering of dust grains, the cooling in shocks, however, is captured self-consistently by the applied chemical network.

The interstellar radiation field (ISRF) is that of \citet{Draine78}, i.e. $G_0$ = 1.7 in Habing units \citep{Habing68}, and its shielding is calculated according to the surrounding column densities of total gas, H$_2$, and CO via the {\sc OpticalDepth} module \citep{Wunsch18} based on the {\sc TreeCol} algorithm \citep{Clark12}. For this purpose, we determine for each cell the visual extinction, $A_\rmn{V, i}$, separately along 48 directions by converting the total gas column density $N_\rmn{H,tot}$ into a visual extinction via \citep{Draine96}
\begin{equation}
 A_\rmn{V, i} = (N_\rmn{H,tot,i} \times 5.348 \times 10^{-22} \rmn{cm}^2) \, \rmn{mag} \, .
\end{equation}
Since the scheme is based on a Healpix tessellation \citep{Gorski11}, all directions are equally weighted. The average local visual extinction in each cell is then obtained as
\begin{equation}
 A_\rmn{V, 3D} = \frac{-1}{\gamma} \textrm{ln} \left( \frac{1}{48} \sum_{i=1}^{48} \textrm{exp}(-\gamma A_\rmn{V, i}) \right) \, ,
 \label{eq:AV}
\end{equation}
with $\gamma$ = 2.5 \citep{Bergin04}. With this definition the local attenuation factor of the ISRF due to dust, which we use for the dissociation reactions of H$_2$ and CO, is then exp(-$\gamma$ $A_\rmn{V, 3D}$) \citep{Glover10}. In addition, similar to Eq.~\ref{eq:AV} for $A_\rmn{V, 3D}$, we calculate the self-shielding of H$_2$ and CO from the H$_2$ and CO column densities, which further reduces the dissociation rates \citep[][but see also Section~2 of \citealt{Glover10}]{Draine96,Lee96}. This approach thus allows us to properly assess the dissociation of H$_2$ and CO in the simulations due to incident UV radiation.

Furthermore, we solve the Poisson equation for self-gravity with a tree-based method \citep{Wunsch18} and include a background potential from the old stellar component in the galactic disc, modeled as an isothermal sheet with \mbox{$\Sigma_\mathrm{star}$ = 30 M$_{\sun}$ pc$^{-2}$} and a scale height of \mbox{100 pc.}

Our setup represents a small section of a stratified galactic disc with solar neighborhood properties and a size of 500 pc $\times$ 500 pc $\times$ $\pm$ 5~kpc. The gas surface density is \mbox{$\Sigma_\mathrm{gas}$ = 10 M$_{\sun}$ pc$^{-2}$} and the initial vertical gas distribution has a Gaussian profile with a scale height of 30 pc and a midplane density of $\rho_0$ =  9 $\times$ $10^{-24}$ g cm$^{-3}$. The gas is initially at rest. Near the disc midplane it has an initial temperature of 4500 K and consists of atomic hydrogen and C$^+$. For the magnetised runs, we initialize a magnetic field along the $x$-direction as
\begin{equation}
 B_{x} = B_{x,0} \sqrt{\rho(z)/\rho_0} \, ,
\end{equation}
where we set the magnetic field in the midplane to \mbox{$B_{x,0}$ = 3 $\mu$G} in accordance with recent observations \citep[e.g.][]{Beck13}.

Up to $t_0$ (see Table~\ref{tab:overview}), we drive turbulence in the disc with supernovae (SNe). Half of the SNe are randomly placed in the $x$-$y$-plane following a Gaussian profile with a scale height of 50 pc in the vertical direction, the other half is placed at density peaks. The SN rate is constant at 15 SNe Myr$^{-1}$, corresponding to the Kennicutt-Schmidt star formation rate surface density for \mbox{$\Sigma_\mathrm{gas}$ = 10 M$_{\sun}$ pc$^{-2}$} \citep{Kennicutt98} and assuming a standard initial mass function \citep{Chabrier01}. For a single SN we inject 10$^{51}$ erg in the form of thermal energy if the Sedov-Taylor radius is resolved with at least 4 grid cells. Otherwise, we heat the gas within the injection region to \mbox{$10^4$ K} and inject the momentum, which the swept-up shell has gained at the end of the Sedov-Taylor phase \citep[see][for details]{Gatto15}.

The base grid resolution is 3.9~pc up to $t_0$. At $t_0$ we stop further SN explosions. We choose different regions in which MCs are about to form. These ``zoom-in'' regions have a rectangular shape with a typical linear extent of about 100 pc. We then continue the simulations for another 1.5~Myr over which we progressively increase the spatial resolution in these zoom-in regions from 3.9~pc to 0.12~pc assuring that the Jeans length is refined with 16 cells \citep[][Table~2]{Seifried17}. In the surroundings we keep the lower resolution of 3.9~pc. Afterwards we continue the simulations with the highest resolution of 0.12 pc in the zoom-in regions.

We consider two purely hydrodynamical (HD) simulations without magnetic fields \citep[runs MC1-HD and MC2-HD, see][ and Table~\ref{tab:overview}]{Seifried17} and two simulations with magnetic fields \citep[MC3-MHD and MC4-MHD, see][]{Seifried19}. For these runs we turn off sink particle formation and stellar feedback inside the clouds. The MCs with and without magnetic fields emerge from different stratified galactic disc simulations. As magnetic fields delay the formation of dense molecular gas \citep{Walch15,Girichidis18}, we start to zoom in at a somewhat later time ($t_0$) for the magnetised runs, such that the cloud masses of a few 10$^{4}$ M$_{\sun}$ are roughly comparable for all four clouds.

Furthermore, in order to investigate the effect of stellar radiative feedback, we rerun MC1-HD and MC2-HD including sink particles and radiative stellar feedback \citep[runs MC1-HD-FB and MC2-HD-FB, see][]{Haid19}. A comparison between these runs allows us to isolate the impact of feedback from that of different initial conditions. Feedback from the massive stars sets in at \mbox{$t$ = 13.8 Myr} and 13.6~Myr for run MC1-HD-FB and MC2-HD-FB, respectively.

In the two feedback runs, sink particles are used to model the formation of stars or star clusters and their subsequent radiative stellar feedback. The sinks form from Jeans-unstable gas once the gas density exceeds a value of 1.1~$\times$~10$^{-20}$ g cm$^{-3}$ and are treated with a 4th-order Hermite predictor-corrector scheme (Dinnbier et al., in prep.). We assure that the cells hosting the sinks are always refined to the highest level of refinement. As time evolves, the sinks accrete gas and form stars. Every 120 M$_{\odot}$ of accreted mass, one massive star between 9 and 120 M$_{\odot}$ is randomly sampled from an initial mass function assuming a slope of -2.3 between 9 and 120~M$_{\sun}$ \citep{Salpeter55}.

Each massive star follows its individual, mass-dependent stellar evolutionary track \citep{Ekstrom12,Gatto17,Peters17} where we follow in detail the amount of photoionizing radiation released by each star \citep{Haid18,Haid19}. The radiative feedback is treated with a backwards ray-tracing algorithm {\sc TreeRay} \citep[][W\"unsch et al., in prep.]{Wunsch18}, which efficiently uses the available octal-tree structure. The radiative transport equation is solved for hydrogen-ionizing EUV radiation assuming the {\it On-the-Spot approximation} with a temperature dependent case B recombination coefficient \citep{Draine11}. The resulting number of hydrogen-ionizing photons and the associated heating rate are processed within the chemical network \citep{Haid18}.

\begin{table}
\caption{Overview of the simulations giving the run name, the starting time of the zoom-in procedure $t_0$, the time $t_\rmn{end}$ up to which the clouds are evolved, whether magnetic fields (B, no B) or radiative feedback (FB, no FB) are included, and the underlying reference.}
\centering
\begin{tabular}{lcccc}
  \hline
 run & $t_0$ (Myr) & $t_\rmn{end}$ (Myr) & run type & Ref. \\
 \hline
 MC1-HD & 11.9 & $t_0$ + 4.0 & no B, no FB & (1) \\
 MC2-HD & 11.9 & $t_0$ + 4.0 & no B, no FB & (1) \\
 MC1-HD-FB & 11.9 & $t_0$ + 4.0 & no B, FB & (2)  \\
 MC2-HD-FB & 11.9 & $t_0$ + 4.0 & no B, FB & (2)  \\
 MC3-MHD & 16.0 & $t_0$ + 5.5 & B, no FB & (3) \\
 MC4-MHD & 16.0 & $t_0$ + 5.5 & B, no FB & (3) \\
 \hline
 \end{tabular}
 \\
(1) \citet{Seifried17}, 
(2) \citet{Haid19}, 
(3) \citet[][in this reference, the runs are denoted as "MC1" and "MC2"]{Seifried19}
 \label{tab:overview}
\end{table}

\section{Results}
\label{sec:results}

\begin{figure*}
 \includegraphics[width=\linewidth]{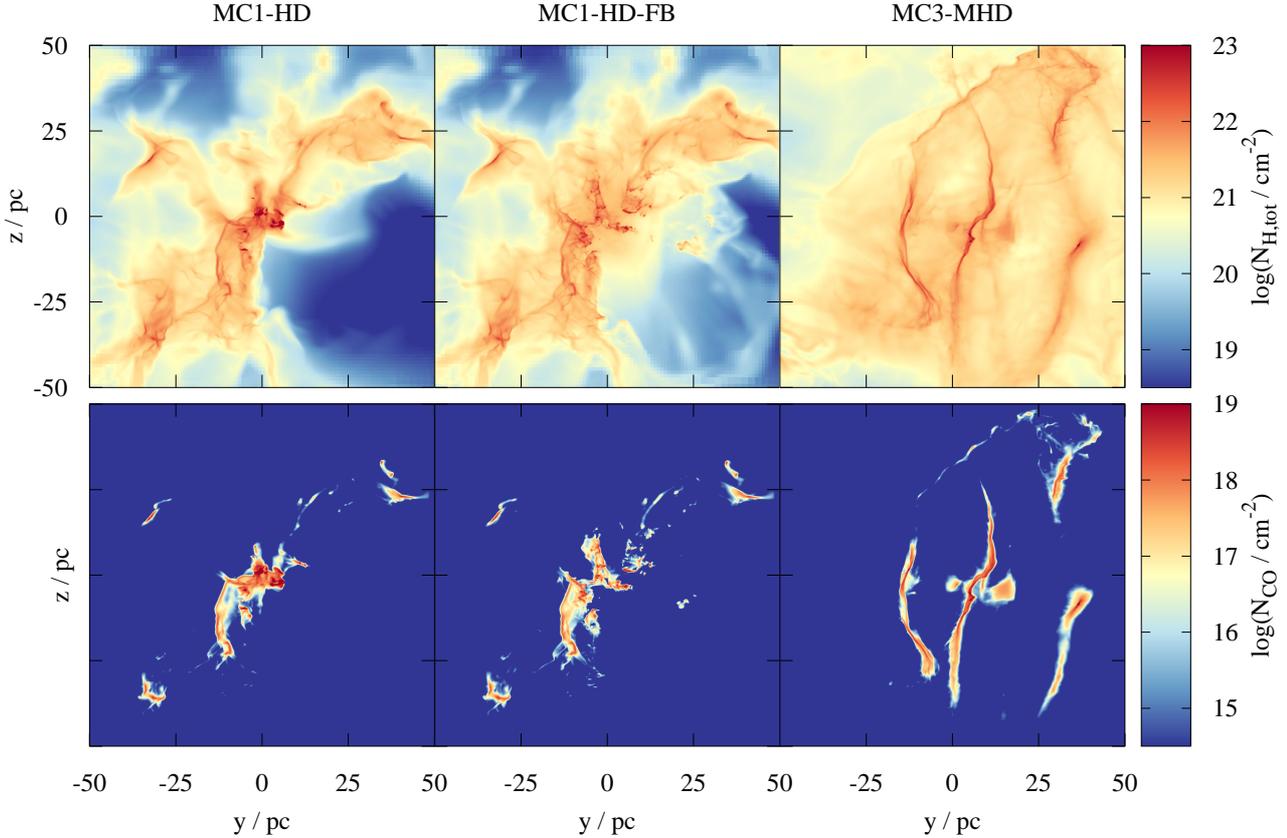}
 \caption{Column density of hydrogen nuclei (top row) and CO (bottom row) of the runs MC1-HD, MC1-HD-FB, and MC3-MHD (from left to right) at \mbox{$t_\rmn{evol}$ = 4 Myr} projected along the $x$-direction. The color bars of the $N_\rmn{H,tot}$ and $N_\rmn{CO}$ maps have the same dynamical range of 4.5 mag, which allows for a direct comparison of both maps: CO shows a significantly more compact distribution than that of the total gas leading to a significant amount of CO-dark gas, in particular in the outer regions of the cloud. Radiative feedback (middle) blows apart the dense structures in the centre of the cloud seen in the corresponding run without feedback (left). The MC with magnetic fields (right) shows a somewhat more diffuse and filamentary structure.}
 \label{fig:column_dens}
\end{figure*}

Throughout the paper we refer to the time elapsed since $t_0$ as \mbox{$t_\rmn{evol}$ = $t$ - $t_0$}. Hence, for the runs MC1-HD-FB and MC2-HD-FB, feedback sets in at $t_\rmn{evol}$ =  1.9 and 1.7 Myr, respectively. In the following we constrain ourselves to the times $t_\rmn{evol}$ = 2, 3, 4 and 5 Myr, with the latest time being considered only for the runs MC3-MHD and MC4-MHD, which evolve more slowly\footnote{Note that assuming a typical turbulent velocity of $\sim$ 5 km s$^{-1}$ \citep[see Fig.~5 of][]{Seifried17} and an extent of $\lesssim$ 50 pc, we obtain characteristic turnover timescales of $\lesssim$ 10 Myr.}. In Fig.~\ref{fig:column_dens} we show the hydrogen nuclei column density and the CO column density of the runs MC1-HD, MC1-HD-FB, and MC3-MHD at $t_\rmn{evol}$ = 4 Myr along the $x$-direction. Overall, CO shows a significantly more compact distribution leading to the problem of CO-dark gas discussed in detail in the following. The typical CO column densities span a range from $\sim$ 10$^{16}$ -- 10$^{19}$ cm$^{-2}$, peaking around 10$^{17}$ cm$^{-2}$. This is comparable with actual observations of e.g. the Taurus molecular cloud~\citep{Goldsmith08}. A more detailed analysis of the CO column density distribution, however, will be presented in a subsequent paper (Borchert et al., in prep.). For further details on the dynamical evolution of the clouds we refer to the references given in Table~\ref{tab:overview}. In the following we mainly focus on their chemical composition.

\subsection{The CO-dark gas fraction in molecular clouds}
\label{sec:dgf_total}

As already visible by eye from Fig.~\ref{fig:column_dens}, the distribution of CO is significantly more compact than the distribution of the total gas. For this reason, we first determine the global mass fraction of CO-dark gas (henceforth DGF) in our simulated MCs using the full 3D information. For this purpose we calculate the ratio of the total CO mass, $M_\rmn{CO}$, to the total H$_2$ mass, $M_\rmn{H_2}$, in the zoom-in region. We correct for the fact that in our simulations the total fractional abundance of carbon with respect to hydrogen nuclei is 1.4 $\times$ 10$^{-4}$, or 2.8 $\times$ 10$^{-4}$ with respect to H$_2$ molecules (under the assumption that hydrogen is completely in its molecular form), i.e.
\begin{equation}
 \textrm{DGF} = 1 - \frac{\frac{M_\rmn{CO}}{28 m_\rmn{p}} \times \frac{1}{2.8 \times 10^{-4}}}{\frac{M_\rmn{H_2}}{2 m_\rmn{p}}} \, , 
 \label{eq:dgf}
\end{equation}
where $m_\rmn{p}$ is the proton mass. Note that this definition describes the fraction of \textit{intrinsically} CO-dark gas, i.e. gas with no CO molecules but H$_2$ molecules, which is only accessible in simulations. It thus differs from the common observational definition of a DGF, which is based on observational sensitivity limits for CO observations \citep[][but see also our Section~\ref{sec:limit}]{Wolfire10}.

At this point we would like to note that in observational literature, the terms ''CO-dark``, ''CO-poor`` and ''CO-faint`` appear to be used interchangeably \citep[e.g.][]{Lada88,Dishoeck92,Grenier05,Bolatto13,Velusamy14}. In this work, with (intrinsically) ''CO-dark`` gas we refer to the actual molecular content (Eq.~\ref{eq:dgf}) and with ''CO-faint`` gas to an observational definition (Section~\ref{sec:limit}).

\begin{figure}
 \includegraphics[width=0.9\linewidth]{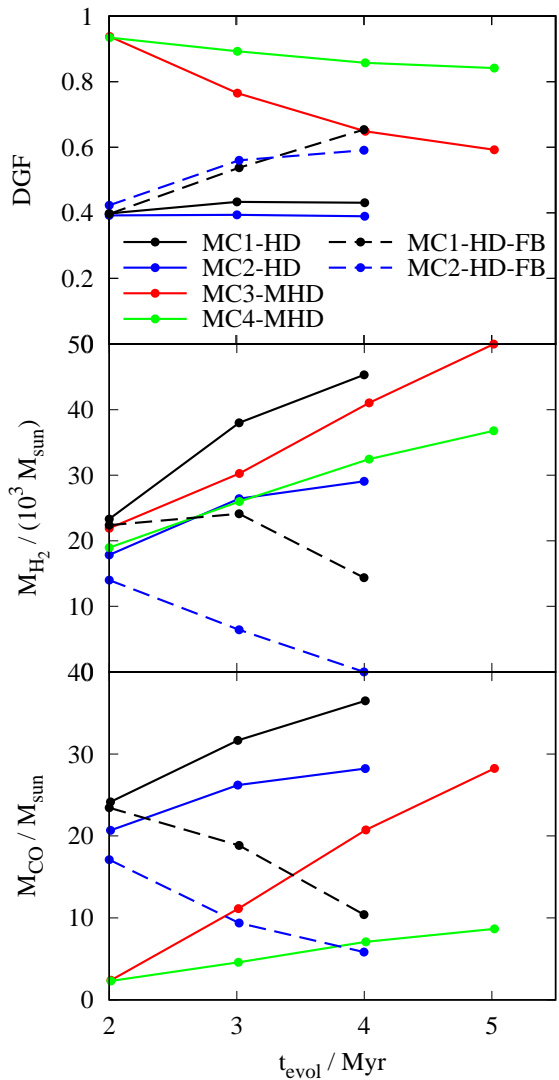}
 \caption{Time evolution of the global DGF (Eq.~\ref{eq:dgf}, top), the total H$_2$ mass (middle) and the total CO mass (bottom) for MCs without feedback (solid lines) and with feedback (dashed lines). For the runs without feedback, the DGF remains roughly constant in the absence of magnetic fields (solid black and blue lines), whereas for runs including magnetic fields it decreases (red and green lines). In both cases the total amount of H$_2$ increases over time. Feedback increases the global DGF and reduces the overall amount of H$_2$ and, even more efficiently, the amount of CO.}
 \label{fig:dgf_total}
\end{figure}
\begin{figure}
  \includegraphics[width=\linewidth]{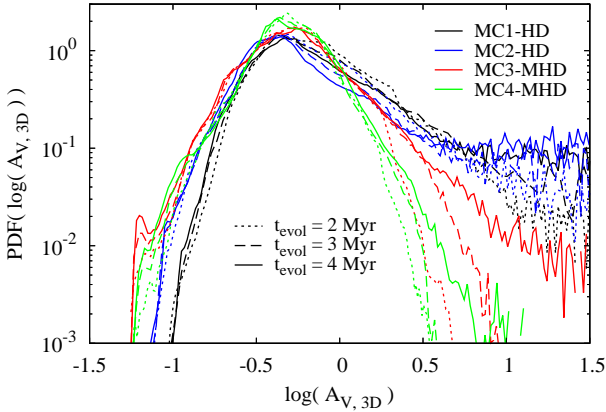}\\
  \caption{Mass-weighted $A_\rmn{V, 3D}$-PDF of the runs with and without magnetic fields at $t_\rmn{evol}$ = 2, 3 and 4 Myr. The MHD runs have less well-shielded gas, resulting in less CO and higher DGFs (Fig.~\ref{fig:dgf_total}).}
  \label{fig:AV_pdf}
\end{figure}

In the top panel of Fig.~\ref{fig:dgf_total} we show the DGF as a function of time. We first focus on the runs without radiative feedback. For the hydrodynamical runs (MC1-HD and MC2-HD, black and blue solid lines), the DGF remains roughly constant over time with values of $\sim$ 0.4. In the presence of magnetic fields (runs MC3-MHD and MC4-MHD, red and green lines), however, the amount of CO-dark gas is initially significantly higher with values around 0.95 at \mbox{$t_\rmn{evol}$ = 2 Myr} and then decreases over time to values of 0.6 and 0.85, respectively, as the clouds become increasingly denser and more CO forms. Interestingly, the evolution of $M_\rmn{H_2}$ (middle panel of Fig.~\ref{fig:dgf_total}) is similar for all four runs. It increases over time with a spread among the simulations of about 10$^4$ M$_{\sun}$, i.e. relative differences of $\sim$ 20\%. Hence, the significantly higher fraction of CO-dark gas for the MHD clouds (about a factor of 2) cannot be attributed to changes in $M_\rmn{H_2}$ but to a lower amount of CO in these runs (bottom panel of Fig.~\ref{fig:dgf_total}). We attribute this lower amount of CO to differences in the structure of the clouds. This becomes already apparent by eye when investigating Fig.~\ref{fig:column_dens}. The MCs with magnetic fields appear to be more diffuse and filamentary than the MCs without magnetic fields. For a fixed $M_\rmn{H_2}$, however, a more filamentary structure would result -- on average -- in lower visual extinctions and thus lower $M_\rmn{CO}$ \citep{Roellig07,Glover10}. For completeness, in Fig.~\ref{fig:density} in Appendix A, we also show the mean H$_2$ and CO densities in the dense gas ($n$ $\geq$ 100 cm$^{-3}$) and the global mass fractions of both species in the zoom-in regions.

At this point, we emphasize that here and in Section~\ref{sec:distribution} we consider the local visual extinction of the ISRF at each point in the cloud, i.e. $A_\rmn{V, 3D}$, which is calculated directly during the simulation via the {\sc OpticalDepth} module \citep[Eq.~\ref{eq:AV} and][]{Wunsch18}. It thus gives an reasonable approximation of the local attenuation of the ISRF, but does not directly correspond to the visual extinction obtained in observations by averaging along the line-of-sight (LOS), which we consider in Sections~\ref{sec:2dmaps} and~\ref{sec:newapproach}.

In Fig.~\ref{fig:AV_pdf}, we plot the mass-weighted probability density function (PDF) of log($A_\rmn{V, 3D}$) for the four different MCs without feedback at $t_\rmn{evol}$ = 2, 3 and 4 Myr, where the mass-weighted PDF of a quantity $x$ is given by
\begin{equation}
 \rmn{PDF}(x) = \frac{\rmn{d}m}{\rmn{d}x} \frac{1}{M_\rmn{tot}} \, .
\end{equation}
As speculated before, the magnetised clouds MC3-MHD and MC4-MHD are more diffuse objects with significantly smaller mass fractions at $A_\rmn{V, 3D}$ $>$ 1 than the clouds without magnetic fields (MC1-HD and MC2-HD). This is also supported by the mean densities of H$_2$ shown in the top left panel of Fig.~\ref{fig:density}. Hence, as CO only starts to form at \mbox{A$_\rmn{V, 3D}$ $>$ 1}, and H$_2$ already at $A_\rmn{V, 3D}$ $\gtrsim$  0.3 \citep[e.g.][]{Roellig07,Glover10,Bisbas19}, this explains the observed differences in the DGF. It also matches our previous findings that magnetic fields significantly hamper the formation of dense, well-shielded molecular gas \citep{Walch15,Girichidis18}. Our findings are, however, in contradiction to the result of 1D calculations of \citet{Wolfire10}, who claim that the amount of CO-dark gas is insensitive to the internal density -- and thus $A_\rmn{V, 3D}$ -- distribution, thus emphasising the need of 3D, MHD simulations. We note that the increase of mass at $A_\rmn{V, 3D}$ $>$ 1 is accompanied with an increase in $M_\rmn{CO}$ (bottom panel of Fig.~\ref{fig:dgf_total}), which is particularly pronounced for MC3-MHD (red lines). We will investigate the dependence of the DGF on the shielding in more detail in the next section confirming the results shown so far.

For runs including feedback (MC1-HD-FB and MC2-HD-FB, dashed lines in Fig.~\ref{fig:dgf_total}), the DGF increases over time from $\sim$ 0.4 to $\sim$ 0.7 as CO is apparently destroyed more efficiently via photodissociation than H$_2$ (compare middle and bottom panel). We attribute this to the fact that (i) radiative feedback from young, massive stars acts where the stars are born, i.e. preferentially in the densest regions of MCs, which are fully molecular, and (ii) the dissociation rate per molecule and per UV photon incorporated in our chemical network is a factor of 3.86 times higher for CO than for H$_2$ \citep{Dishoeck88,Roellig07}. Hence, MCs actively forming massive stars appear to have a higher amount of CO-dark gas than their quiescent counterparts.

The overall high values of the DGF and its large spread agree well with recent observations showing DGFs in MCs of 30\% and more \citep{Grenier05,Lee12,Planck15}. Our results are also in agreement with theoretical results \citep{Wolfire10,Smith14,Gong18,Li18}, although the definition used by these authors does not exactly match the DGF as defined in Eq.~\ref{eq:dgf} (see Section~\ref{sec:limit} for details).

\subsection{The distribution of CO-dark and CO-bright gas}
\label{sec:distribution}

\subsubsection{Dependence on the local $A_\rmn{V, 3D}$}
\label{sec:AV3D}

\begin{figure*}
 \includegraphics[width=\textwidth]{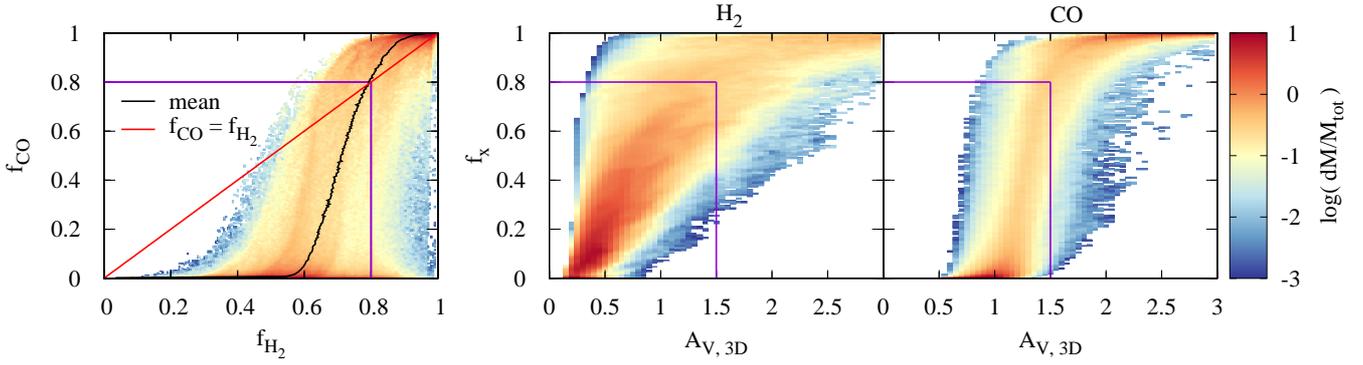}\\
 \caption{Left: Mass-weighted $f_\rmn{H_2}$-$f_\rmn{CO}$-phase diagram of MC1-HD at $t_\rmn{evol}$ = 3 Myr. The black solid line shows the mean of the distribution, the red line the 1:1 line. Both mass fractions become roughly comparable above $f_x$ $\simeq$ 0.8 $\pm$ 0.1 (here the subscript $x$ stands for H$_2$ and CO, respectively). Middle and right: Mass-weighted phase diagram of $A_\rmn{V, 3D}$ and $f_\rmn{H_2}$ (middle) and $f_\rmn{CO}$ (right). In order to guide the readers eye, we draw a violet line at \mbox{$A_\rmn{V, 3D}$ = 1.5}, where both mass fractions become comparable. This value of $A_\rmn{V, 3D}$ is in reasonable agreement with the drop of the DGF to zero found in Fig.~\ref{fig:dgf}.}
 \label{fig:phasediag} 
\end{figure*}

As shown before, the differences in the DGF by a factor of $\lesssim$~2 between runs with and without magnetic fields (Fig.~\ref{fig:dgf_total}) can be attributed to different morphologies of the clouds, resulting in less mass at $A_\rmn{V, 3D}$ $>$ 1 for the magnetised clouds (Fig.~\ref{fig:AV_pdf}). For this reason, we next consider the distribution of H$_2$ and CO relative to each other and with respect to the local visual extinction, $A_\rmn{V, 3D}$. For this purpose, we determine the mass fractions of H$_2$ and CO
\begin{equation}
 f_\rmn{H_2} = \frac{m_\rmn{H_2,cell}}{m_\rmn{H,cell,tot}} \; \; \; \textrm{and} \; \; \; f_\rmn{CO} = \frac{\frac{12}{28} \, m_\rmn{CO,cell} }{m_\rmn{C,cell,tot}} \, ,
 \label{eq:fraction}
\end{equation}
where $m_\rmn{H,cell,tot}$ and $m_\rmn{C,cell,tot}$ are the total mass of hydrogen and carbon available in the considered cell, $m_\rmn{H_2,cell}$ and $m_\rmn{CO,cell}$ the mass of all H$_2$ and CO molecules, respectively, and the factor $\frac{12}{28}$ corrects for the mass of oxygen.

In Fig.~\ref{fig:phasediag} we show the results for run MC1-HD at $t_\rmn{evol}$ = 3 Myr. The qualitative behaviour also holds for all other runs and times. As can be seen from the left panel, significant amounts of CO are only formed once $\sim$ 50\% of the hydrogen is in molecular form. Both mass fractions become roughly comparable above \mbox{$f_x$ $\simeq$ 0.8 $\pm$ 0.1} (where the subscript $x$ stands for H$_2$ and CO, respectively). This value is a rather rough estimate, which also depends on the considered simulation and increases over time ranging from $\sim$ 0.5 -- 0.8.

Considering the middle and right panel of Fig.~\ref{fig:phasediag}, we find that for both H$_2$ and CO \mbox{$f_x$ = 0.8} is approximately reached at \mbox{$A_\rmn{V, 3D}$ $\simeq$ 1.5} (see violet lines), around which we expect the DGF to drop to zero. At lower $A_\rmn{V, 3D}$ of a few 0.1, $f_\rmn{H_2}$ is as high as 0.1 -- 0.2 (middle panel), i.e. noticeably higher than $f_\rmn{CO}$, which remains close to zero in this range (right panel) and starts to rise around \mbox{$A_\rmn{V, 3D}$ $\simeq$ 1} in good agreement with detailed chemical PDR models \citep[e.g.][see also \citealt{Gong18} for similar results in 3D-MHD simulations]{Roellig07,Glover10}. At even lower values of $A_\rmn{V, 3D}$ ($\lesssim$ 0.1), neither H$_2$ nor CO are present. We note that sometimes $f_\rmn{CO}$ can be slightly higher than $f_\rmn{H_2}$, which can be attributed to the short formation time of CO, once a sufficient amount of H$_2$ is present \citep[see Eq. 9 in][but also \citealt{Nelson97,Glover10}]{Seifried17}.

\begin{figure*}
 \includegraphics[width=\textwidth]{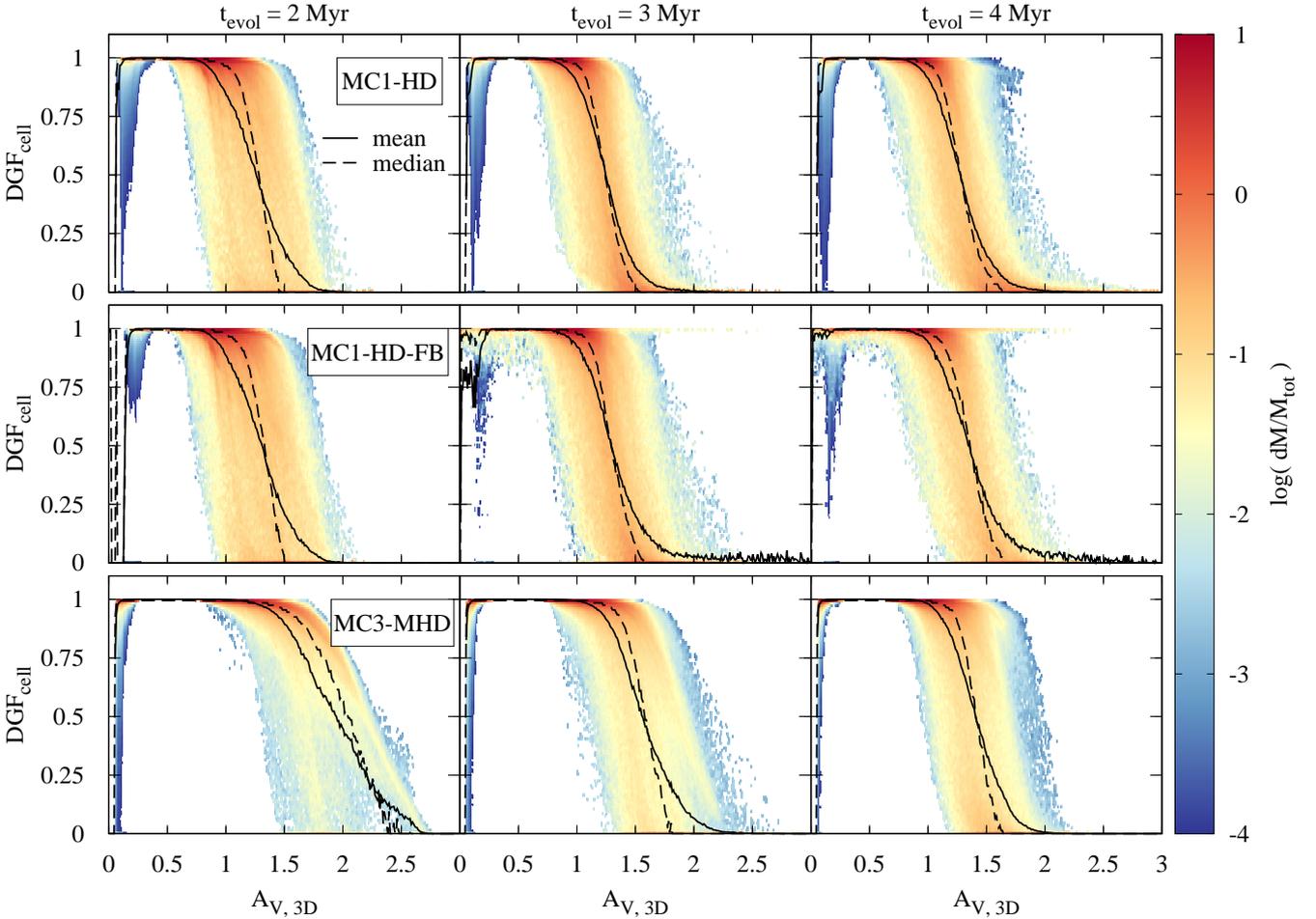}
 \caption{Mass-weighted $A_\rmn{V, 3D}$-DGF$_\rmn{cell}$-phase diagram for different times (left to right) for MC1-HD (top), MC1-HD-FB (middle) and MC3-MHD (bottom). DGF$_\rmn{cell}$ represents the amount of CO-dark gas in each cell (Eq.~\ref{eq:dgf_f}), i.e. for a value of 1 no CO is present. The black solid and dashed lines denote the mean and median of the distribution. The overall shape of the phase diagrams is similar for all times and runs considered, with a drop of DGF$_\rmn{cell}$ around $A_\rmn{V, 3D}$ = 1.5.}
 \label{fig:dgf} 
\end{figure*}
Next, we investigate how the DGF of individual cells depends on $A_\rmn{V, 3D}$. The DGF in a cell is related to the mass fractions of H$_2$ and CO (Eq.~\ref{eq:fraction}) as
\begin{equation}
 \textrm{DGF}_\rmn{cell} = 1 - \frac{f_\rmn{CO}}{f_\rmn{H_2}} \, .
 \label{eq:dgf_f}
\end{equation}
In Fig.~\ref{fig:dgf} we show the $A_\rmn{V, 3D}$-DGF$_\rmn{cell}$-phase diagram for the clouds MC1-HD, MC1-HD-FB and MC3-MHD at $t_\rmn{evol}$ = 2, 3 and 4 Myr. We emphasize that for the remaining runs the phase diagrams are similar. The general shape of the distribution changes only moderately with time and for the different runs. Once molecular hydrogen starts to form ($A_\rmn{V, 3D}$~$\gtrsim$~0.1), DGF$_\rmn{cell}$ quickly rises to $\sim$~1. It remains high until \mbox{$A_\rmn{V, 3D}$ $\simeq$ 1} and then drops to almost zero around $A_\rmn{V, 3D}$ = 1.5 -- 2 as indicated by the black solid and dashed lines which denote the mean and median of the distribution. This is in good agreement with our findings that the mass fractions of H$_2$ and CO become comparable at \mbox{$A_\rmn{V, 3D}$ $\simeq$ 1.5} (Fig.~\ref{fig:phasediag}). Furthermore, the observed drop of DGF$_\rmn{cell}$ around $A_\rmn{V, 3D}$ = 1.5 is also in good agreement with the findings of \citet{Xu16} in the Taurus molecular cloud.

For the runs with stellar feedback (middle row of Fig.~\ref{fig:dgf}), somewhat more CO-dark gas appears at $A_\rmn{V, 3D}$ $>$ 1.5 at later stages (seen as a horizontal stripe in the phase diagram). The overall shape, however, remains almost unchanged. For MC3-MHD (bottom row), initially (\mbox{$t_\rmn{evol}$ $\leq$ 3 Myr}), the drop of DGF$_\rmn{cell}$ seems to appear at slightly higher $A_\rmn{V, 3D}$. As at this evolutionary stage the structure of the cloud is significantly more diffuse (Fig.~\ref{fig:AV_pdf}), we speculate that also the self-shielding due to H$_2$ and CO, which prevents CO from being dissociated, is reduced. Hence, CO forms at higher $A_\rmn{V, 3D}$ than for the more compact hydrodynamical clouds. However, towards later stages, the phase-diagrams approach those from the runs without magnetic fields. Similar results are also found for MC4-MHD (not shown here).

\begin{figure}
 \includegraphics[width=0.9\linewidth]{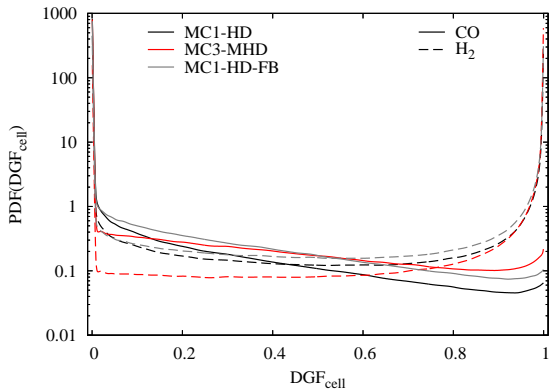}
 \caption{CO (solid line) and H$_2$ (dashed line) mass-weighted PDF of DGF$_\rmn{cell}$ for the runs MC1-HD, MC3-MHD and MC1-HD-FB at $t_\rmn{evol}$ = 3~Myr. The majority of CO sits in cells with a low DGF$_\rmn{cell}$, whereas H$_2$ can also be found in cells with a high DGF$_\rmn{cell}$.}
 \label{fig:pdf_dgf} 
\end{figure}
Integrating the plots in Fig.~\ref{fig:dgf} along the $x$-axis gives the amount of gas at a given DGF$_\rmn{cell}$. In Fig.~\ref{fig:pdf_dgf} we show the PDF of DGF$_\rmn{cell}$ weighted by the CO and H$_2$ mass, respectively. For the sake of readability, we only show the lines for the runs MC1-HD, MC3-MHD, and MC1-HD-FB, but note that the remaining runs are qualitatively and quantitatively very similar. As can be seen, the vast majority (note the logarithmic scaling of the $y$-axis) of CO sits in cells with a low DGF$_\rmn{cell}$. For H$_2$, however, significant amounts of gas can be found close to DGF$_\rmn{cell}$ = 1 and DGF$_\rmn{cell}$ = 0, which is in good agreement with the partly high global DGF (Fig.~\ref{fig:dgf_total}).

To summarize, our results indicate that -- despite significant variations in the absolute amount --  CO-dark gas is mainly present in gas with visual extinctions 0.2 -- 0.3 $<$ $A_\rmn{V, 3D}$ $<$ 1 -- 1.5, independent of the presence or absence of magnetic fields or stellar feedback. Above $A_\rmn{V, 3D}$ $\simeq$ 1.5, the gas is mainly CO-bright, which happens once about 50 -- 80\% of both hydrogen and carbon are in molecular form.

\subsubsection{Dependence on density and temperature}

\begin{figure*}
 \includegraphics[width=\textwidth]{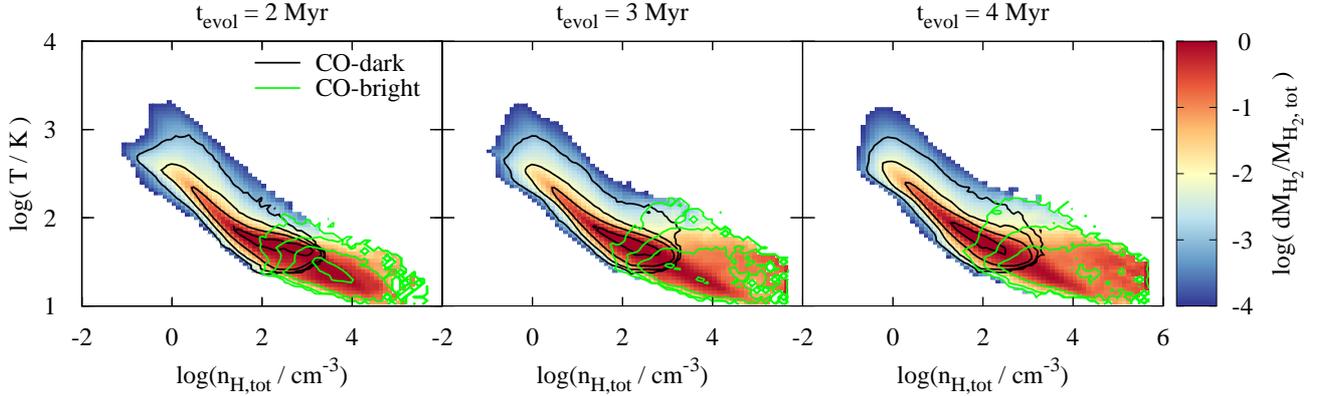}\\
 \caption{Time evolution (from left to right) of the H$_2$-mass-weighted $n_\rmn{H,tot}$-$T$-phase diagram for run MC1-HD without feedback (see Fig.~\ref{fig:temp_dens_diag_FB_MHD} for other runs). The colour code shows the complete H$_2$-mass weighted $n_\rmn{H,tot}$-T-phase diagram, while the black (green) contours show the CO-dark (CO-bright) H$_2$ gas according to the definition that CO-bright gas has a DGF$_\rmn{cell}$ $>$ 0.5. The contour intervals are in steps of 1 from log(d$M_\rmn{H_2}/M_\rmn{H_2,tot}$) = -3 to 0.}
 \label{fig:temp_dens_diag}
\end{figure*}
Next, in Fig.~\ref{fig:temp_dens_diag} we investigate the distribution of CO-dark and CO-bright gas in the density ($n_\rmn{H,tot}$) -- temperature (T) phase space for a run without feedback (MC1-HD). The findings discussed in the following are also representative for the remaining runs with feedback and magnetic fields (see Fig.~\ref{fig:temp_dens_diag_FB_MHD} in the Appendix). We define the CO-dark gas (black contours) as all H$_2$ gas in cells with DGF$_\rmn{cell}$ $>$ 0.5 (Eq.~\ref{eq:dgf_f}) and the CO-bright gas (green contours) as all H$_2$ gas in cells with DGF$_\rmn{cell}$ $\leq$ 0.5.

The bulge of CO-bright gas sits at \mbox{$n_\rmn{H,tot}$ $\gtrsim$ 300 cm$^{-3}$} and temperatures below \mbox{$\sim$ 50 K}, although in particular for the runs with feedback (top panel of Fig.~\ref{fig:temp_dens_diag_FB_MHD}) some CO-bright gas can be found at temperatures up to a few 100 K. The bulge of CO-dark gas, however, occurs at densities of \mbox{10 cm$^{-3}$} $\lesssim$ $n_\rmn{H,tot}$ $\lesssim$ \mbox{10$^{3}$ cm$^{-3}$} and temperatures of a few \mbox{$10$ K} $\lesssim$ $T$ $\lesssim$ a few \mbox{100 K}. Our results are thus in rough agreement with the findings of \cite{Glover16b}, who find CO-dark gas to reside at temperatures above $\sim$ 30~K.

There is, however, a substantial overlap of CO-dark and -bright gas in the $n_\rmn{H,tot}$-$T$-plane (see also Section~\ref{sec:2dmaps}). Moreover, radiative feedback (top panel of Fig.~\ref{fig:temp_dens_diag_FB_MHD}) even further extends the region in the $n_\rmn{H,tot}$-$T$-parameter space, in which CO-dark gas is found. We speculate that this broad distribution of CO-dark gas might complicate its identification in actual observations. This is supported by recent theoretical studies showing the necessity of observing various lines like [OI], [CI] and [CII] to capture the entire CO-dark gas component of MCs \citep{Glover16b,Franeck18,Li18,Clark19}. In this context, we note that other tracers like ArH$^+$, HF, or HCl have also been suggested to differentiate between the atomic and molecular (hydrogen) phase in MCs \citep{Schilke95,Schilke14,Neufeld97,Neufeld05,Neufeld16}, which might allow for a more accurate estimate of the H$_2$ content of MCs.

\subsection{The DGF in 2D maps}
\label{sec:2dmaps}

\begin{figure*}
 \includegraphics[width=\textwidth]{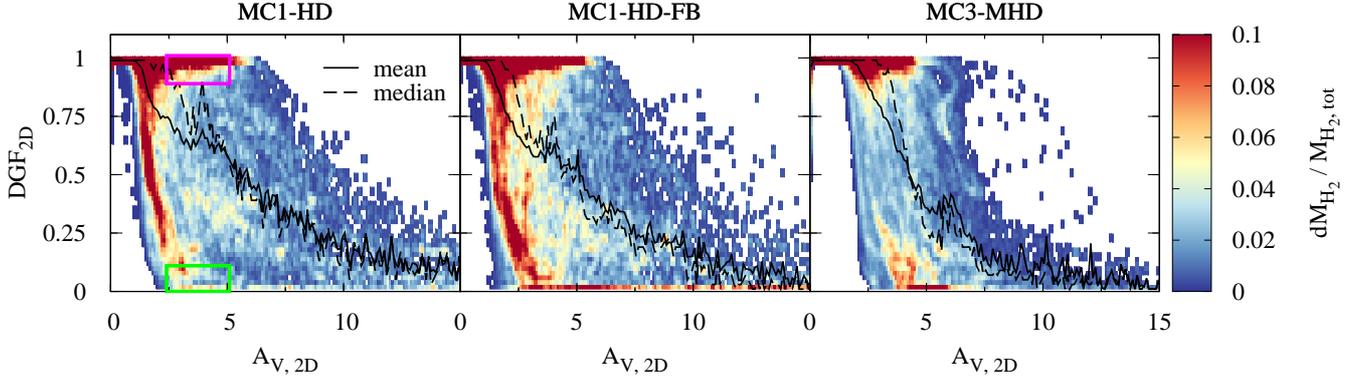} 
 \caption{Mass-weighted $A_\rmn{V, 2D}$-DGF$_\rmn{2D}$-phase diagrams obtained from the 2D, LOS-integrated maps along the $y$-direction of the runs MC1-HD, MC1-HD-FB, and MC3-MHD (from left to right) at \mbox{$t_\rmn{evol}$ = 3 Myr}. Significant fractions of CO-dark gas can be found up to $A_\rmn{V, 2D}$ of $\sim$ 5. In the range of $A_\rmn{V, 2D}$ $\simeq$ 2.5 -- 5 both CO-dark and CO-bright gas is present, which relates to the overlap of CO-dark and -bright gas in the $\rho$ - $T$ - plane (Fig.~\ref{fig:temp_dens_diag}). The two boxes show the region in phase space which are further analysed in Fig.~\ref{fig:AV_LOS}.}
 \label{fig:2DDGF}
\end{figure*}

So far we have based our analysis on the local value of the visual extinction, $A_\rmn{V, 3D}$. This value, however, is not accessible in observations, where only a LOS-integrated column density and the corresponding (2-dimensional) visual extinction are accessible, e.g. via dust extinction measurements \citep[e.g.][]{Lombardi01}. Hence, in order to allow for a better comparison with actual observations, we integrate along the $x$-, $y$-, and $z$-direction of each zoom-in region to obtain maps of the projected total hydrogen column density, $N_\rmn{H, tot}$, and convert it to a visual extinction, $A_\rmn{V, 2D}$, via
\begin{equation}
 A_\rmn{V, 2D} = (N_\rmn{H, tot} \times 5.348 \times 10^{-22} \, \rmn{cm^{2}}) \, \rmn{mag}
 \label{eq:AV2D}
\end{equation}
\citep{Bohlin78,Draine96}. In addition, we calculate the projected DGF for each pixel in the map similar to Eq.~\ref{eq:dgf}:
\begin{equation}
 \textrm{DGF}_\rmn{2D} =   1 - \frac{N_\rmn{CO} \times \frac{1}{2.8 \times 10^{-4}}}{N_\rmn{H_2}} \, ,
\end{equation}
where $N_\rmn{H_2}$ and $N_\rmn{CO}$ are the surface densities of H$_2$ and CO in each pixel. We use a pixel size of 0.12 pc identical to the maximum resolution of the simulations. In Fig.~\ref{fig:2DDGF} we show the resulting H$_2$-mass-weighted $A_\rmn{V, 2D}$-DGF$_\rmn{2D}$-phase diagram obtained from the 2D maps integrated along the $y$-direction for the runs MC1-HD, MC1-HD-FB, and MC3-MHD at \mbox{$t_\rmn{evol}$ = 3 Myr}. Similar results are obtained for the remaining runs, times and directions.

In contrast to the results for $A_\rmn{V, 3D}$ (Fig.~\ref{fig:dgf}) and observational results of \citet{Xu16}, DGF$_\rmn{2D}$ drops to zero at somewhat higher values of $A_\rmn{V, 2D}$ $\simeq$ 2 -- 4. In addition, we find a significant amount of CO-dark gas up to $A_\rmn{V, 2D}$ $\simeq$ 5, where one would not expect it \citep[see Fig.~\ref{fig:dgf}, but also][]{Roellig07,Glover10}. Furthermore, there appears to be a broad distribution of DGF$_\rmn{2D}$ in the range of $A_\rmn{V, 2D}$ $\simeq$ 2.5 -- 5, with both CO-dark and CO-bright gas being present. This demonstrates that, under certain circumstances, $A_\rmn{V, 2D}$  -- which is an average quantity -- can give only little insight about the actual conditions along the entire LOS. This is, however, not surprising given the overlap of CO-dark and -bright gas in the $\rho$ - $T$ - plane found in Fig.~\ref{fig:temp_dens_diag}.

The appearance of CO-dark gas at $A_\rmn{V, 2D}$ $\simeq$ 5 could be understood when considering the average density along the LOS of such a pixel. Assuming a typical length of the LOS of $\sim$ 50 pc, with Eq.~\ref{eq:AV2D} we obtain a total hydrogen column density of $\sim$ 10$^{22}$ cm$^{-2}$ and a volume density of about 60 cm$^{-3}$. Using the relation between the density and $A_\rmn{V, 3D}$ found in our simulations \citep[see Fig.~11 in][]{Seifried17}, such a density corresponds to a typical \mbox{$A_\rmn{V, 3D}$ $\leq$ 1}. Hence, under the assumption that the gas is uniformly distributed along the LOS, we expect CO to not have formed yet, and thus DGF$_\rmn{2D}$ to be close to 1 at $A_\rmn{V, 2D} \simeq 5$.

However, the assumption of a uniform density distribution along the LOS is clearly an oversimplification. Hence, in order to fully understand the reason for the broad distribution of DGF$_\rmn{2D}$ around $A_\rmn{V, 2D}$ $\simeq$ 2.5 -- 5, we investigate the distribution of the local visual extinction $A_\rmn{V, 3D}$ and the density along the LOS of all pixels with
\begin{enumerate}
 \item 2.5 $<$ $A_\rmn{V, 2D}$ < 5 and DGF$_\rmn{2D}$ > 0.9, i.e. CO-dark gas (magenta box in the left panel of Fig.~\ref{fig:2DDGF}) and
 \item 2.5 $<$ $A_\rmn{V, 2D}$ < 5 and DGF$_\rmn{2D}$ < 0.1, i.e. CO-bright gas (green box).
\end{enumerate}

\begin{figure*}
 \includegraphics[width=0.48\textwidth]{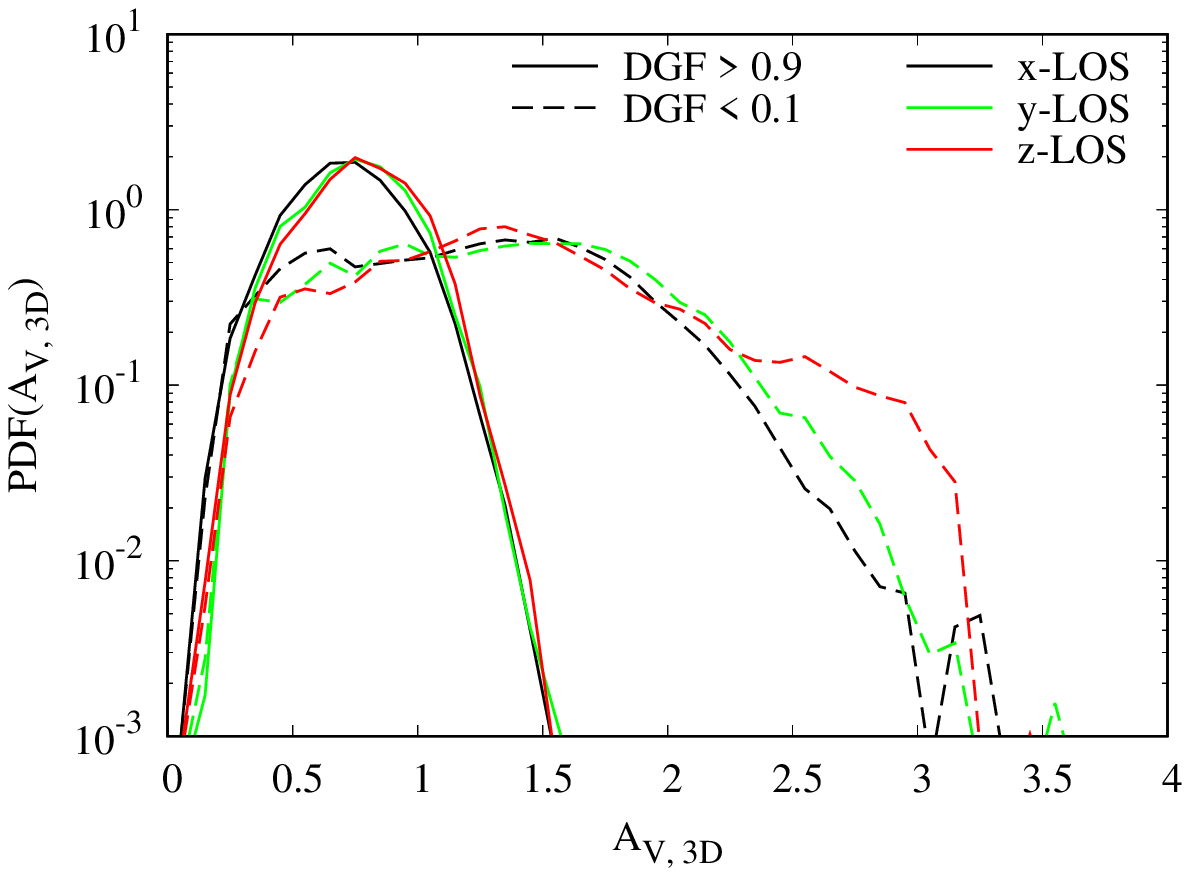} 
 \includegraphics[width=0.48\textwidth]{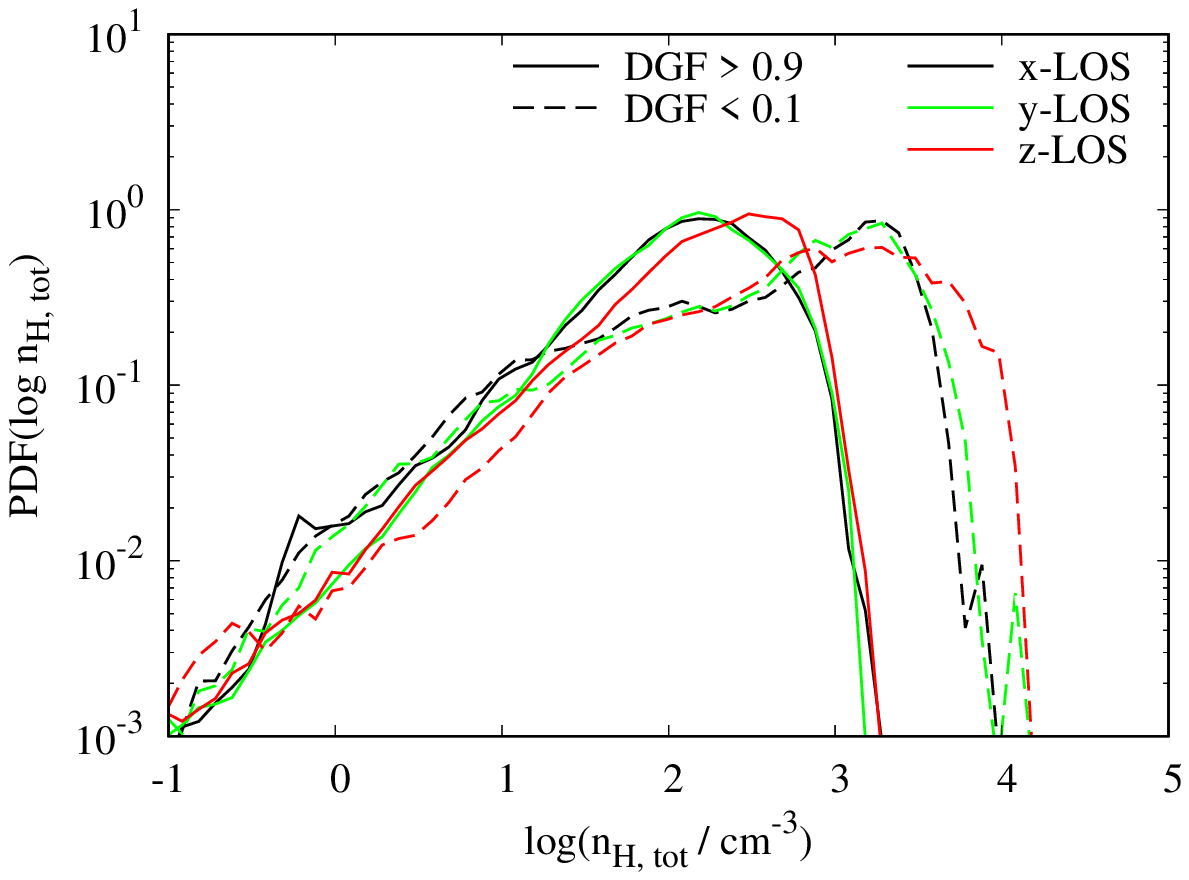}
 \caption{Mass-weighted $A_\rmn{V, 3D}$-PDF (left) and density PDF (right) for CO-dark gas (solid lines, pixels in the magenta box in Fig.~\ref{fig:2DDGF}) and CO-bright gas (dashed lines, pixels in the green box in Fig.~\ref{fig:2DDGF}) of MC1-HD at $t_\rmn{evol}$ = 3 Myr. The broad distribution for DGF$_\rmn{2D}$ found in Fig.~\ref{fig:2DDGF} at $A_\rmn{V, 2D}$ = 2.5 -- 5 can be attributed to different density and thus $A_\rmn{V, 3D}$-distributions along the LOS.}
 \label{fig:AV_LOS}
\end{figure*}
We show the mass-weighted \mbox{$A_\rmn{V, 3D}$-PDF} and \mbox{$n_\rmn{H, tot}$-PDF} of both subsets (i) and (ii) for MC1-HD at \mbox{$t_\rmn{evol}$ = 3 Myr} in Fig.~\ref{fig:AV_LOS}; for the other runs and times we obtain qualitatively and quantitatively similar results. We find no differences in the \mbox{$A_\rmn{V, 3D}$-PDF} for the three directions considered (left panel). There is, however, a clear difference in the distributions for CO-dark and CO-bright gas. CO-dark gas (solid lines) shows a much more narrow $A_\rmn{V, 3D}$-distribution which peaks at \mbox{$A_\rmn{V, 3D}$ $\sim$ 0.7} and reaches a maximum of \mbox{$A_\rmn{V, 3D}$ $\simeq$ 1.5}. For CO-bright gas (dashed lines), however, the distribution is much more wide-spread and shifted to higher $A_\rmn{V, 3D}$ with the peak occurring around $\sim$~1.5.

The density PDF (right panel) shows a corresponding behaviour, which is not surprising given the tight relation between $A_\rmn{V, 3D}$ and the density found in our simulations \citep[Fig.~11 in][]{Seifried17}. For CO-dark gas, the PDF peaks around a value of \mbox{$n_\rmn{H, tot}$ $\sim$ 100 cm$^{-3}$}, which is comparable to the average density obtained under the assumption of a uniform gas distribution along the LOS (see above). For CO-bright gas, however, the PDF peaks at 20 -- 30 times higher densities. Altogether, we can thus attribute the broad distribution of DGF$_\rmn{2D}$ found in Fig.~\ref{fig:2DDGF} to pixels with different density distributions along the LOS: For CO-dark gas at $A_\rmn{V, 2D}$ $\simeq$ 2.5 -- 5, we have a rather uniform density -- and thus $A_\rmn{V, 3D}$ -- distribution with $A_\rmn{V, 3D}$ $\leq$ 1 and hence very little CO, whereas H$_2$ is present already. For the CO-bright gas, however, we have regions with strong density contrasts and locally well-shielded gas \mbox{($A_\rmn{V, 3D}$ $\geq$ 1.5)}, causing CO to form. This is in excellent agreement with complementary theoretical \citep{Levrier12} and observational works \citep{Busch19}, which both find that the abundance of CO is increased by local density enhancements along the LOS.

To summarise, our results indicate that the visual extinction inferred from a LOS-averaging process ($A_\rmn{V, 2D}$) as naturally done in observations, is a partly misleading quantity to assess the CO content along the LOS. It should therefore be considered with caution and be complemented with actual CO observations. Furthermore, for a given $A_\rmn{V, 2D}$, the actual $A_\rmn{V, 3D}$-distribution can be relatively broad and show significant qualitative differences for different pixels in agreement with findings of \citet[][their Fig.~12]{Clark14}. We emphasize that recent observations of M17 and Monoceros R2 with the SOFIA telescope also indicate significant emission in [CII] at $A_\rmn{V, 2D}$ $\sim$ 8 -- i.e. implying a high DGF -- in good agreement with our findings (Guevara et al. in prep.)

\subsection{CO observations}
\label{sec:CO}

Next, we investigate to which extent the actual H$_2$ mass, $M_\rmn{H_2}$, can be obtained from $^{12}$CO(1-0) line observations\footnote{In the following we drop the superscript ``12``.}. For this purpose, we use the freely available radiative transfer code RADMC-3D \citep{Dullemond12} to produce synthetic CO(1-0) line emission maps of our MCs along the $x$-, $y$-, and $z$-direction at the same resolution as the simulation data, i.e. 0.12 pc. We use the Large Velocity Gradient method to calculate the level population and the resulting intensity of the CO(1-0) line transitions. The molecular data, e.g. the Einstein coefficients, are taken from the Leiden Atomic and Molecular database \citep{Schoier05}. The line emission maps cover a velocity range of $\pm$ 20 km s$^{-1}$, which guarantees that all emission is captured properly. The channel width is 200 m s$^{-1}$, which results in 201 channels. We show the CO-spectra of the various runs in Fig.~\ref{fig:CO-spectra} in the Appendix. Further discussion of the spectra, however, is beyond the scope of this paper and will be postponed to a subsequent publication (N\"urnberger et al., in prep.).

\subsubsection{The sensitivity of CO observations}
\label{sec:limit}

As a first step we define the fraction of H$_2$ which is in regions with CO emission below the observational sensitivity limit, following the definition of \citet{Wolfire10} in the notation given by \citet[][Eq.~4]{Smith14}:
\begin{equation}
 \Delta f_\rmn{H_2}(x) = \frac{M_\rmn{H_2}^x}{M_\rmn{H_2}^\rmn{CO} + M_\rmn{H_2}^x} = \frac{M_\rmn{H_2}^x}{M_\rmn{H_2}} \, .
 \label{eq:dgf_CO}
\end{equation}
Here, $M_\rmn{H_2}^x$ is the mass of all H$_2$ gas in pixels which have a CO(1-0) intensity of $I_\rmn{CO} \leq x$ and $M_\rmn{H_2}^\rmn{CO}$ all H$_2$ gas in pixels with $I_\rmn{CO} > x$ such that $M_\rmn{H_2}^\rmn{CO} + M_\rmn{H_2}^x = M_\rmn{H_2}$. I.e., assuming an observational sensitivity limit of $I_\rmn{CO} = x$, the fraction $\Delta f_\rmn{H_2}(x)$ of H$_2$ cannot be traced via CO in the observation regardless of how accurately $M_\rmn{H_2}^\rmn{CO}$ can be determined. This CO-faint H$_2$ gas\footnote{We note that $\Delta f_\rmn{H_2}$ has also been denoted as a dark-gas fraction \citep{Wolfire10,Levrier12,Smith14,Gong18,Li18}. However, as stated before, while the DGF defined here via Eq.~\ref{eq:dgf} is a quantity which requires knowledge about the actual CO abundance, which is only accessible via simulation data and does not depend on any threshold, $\Delta f_\rmn{H_2}$ relies on the observable CO intensity and thus depends on the sensitivity limit of the observation.} thus amplifies the problem of \textit{intrinsically} CO-dark gas, i.e. gas which has no CO molecules but H$_2$ molecules (Section~\ref{sec:dgf_total}). We again note that the latter problem exists for the H$_2$ gas inside \textit{and} outside the observable region, i.e. for $M_\rmn{H_2}^\rmn{CO}$ and $M_\rmn{H_2}^x$.

\begin{figure}
\includegraphics[width=\linewidth]{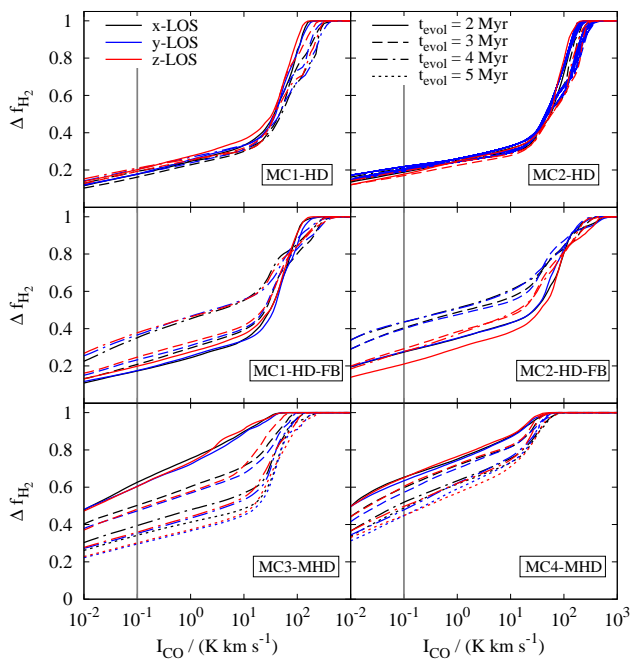}
\caption{Fraction of the total H$_2$ mass, which is in regions with CO emission below an observational sensitivity limit $I_\rmn{CO}$. It is assumed that the amount of H$_2$ above an intensity $I_\rmn{CO}$ can be accurately determined, i.e. no complications due to intrinsically CO-dark gas occur (Section~\ref{sec:dgf_total}). With a typical sensitivity limit of 0.1 K km s$^{-1}$ (indicated with grey lines) a significant amount of H$_2$ is CO-faint, which further amplifies the problem of intrinsically CO-dark gas.}
\label{fig:DGF_CO}
\end{figure}
As shown in Fig.~\ref{fig:DGF_CO}, $\Delta f_\rmn{H_2}$ does only weakly depend on the considered direction. Furthermore, the qualitative shape of the curves seems to be similar for all six MCs: Below a threshold value of $I_\rmn{CO}$ $\simeq$ a few 10 K km s$^{-1}$, the value of $\Delta f_\rmn{H_2}$ decreases steadily with decreasing threshold $I_\rmn{CO}$, which is in agreement with recent observational results of \citet{Donate17}. At $I_\rmn{CO}$ $\simeq$ a few \mbox{10 K km s$^{-1}$}, all curves rise quickly until they reach unity at a few \mbox{100 K km s$^{-1}$} \citep[see also][]{Smith14,Li18}.

However, there are significant differences in the absolute values of $\Delta f_\rmn{H_2}$ for the different MCs. Assuming a typical sensitivity limit of 0.1 K km s$^{-1}$ of recent CO(1-0) observations of nearby MCs \citep[e.g.][indicated by the grey vertical lines in Fig.~\ref{fig:DGF_CO}]{Nieten06,Pineda10,Smith12,Ripple13,Leroy16}, we obtain $\Delta f_\rmn{H_2}(x)$ $\simeq$ 15 -- 65\%. For the runs without feedback, $\Delta f_\rmn{H_2}$ drops over time as more and more CO forms and the intensity in the individual pixels increases. Contrary to that, for the runs with feedback, $\Delta f_\rmn{H_2}$ increases over time as CO gets destroyed by the radiation released from the forming stars. Furthermore, when comparing Fig.~\ref{fig:dgf_total} and~\ref{fig:DGF_CO}, we find a positive correlation between the DGF and  $\Delta f_\rmn{H_2}$.

As the curves in Fig.~\ref{fig:DGF_CO} are rather shallow below  \mbox{$I$ = 10 K km s$^{-1}$}, $\Delta f_\rmn{H_2}$ is not very sensitive on the chosen intensity threshold. Increasing the sensitivity limit to  \mbox{1 K km s$^{-1}$} typical for larger-scale CO surveys \citep[e.g.][and references therein]{Dame01} increases $\Delta f_\rmn{H_2}$ only to values of 20 -- 75\%. Conversely, for a (hypothetical) 10 times lower CO(1-0) sensitivity limit of \mbox{0.01 K km s$^{-1}$}, the fraction of CO-faint H$_2$ gas would be only marginally reduced to $\sim$ 10 -- 50\%. We note that similar values of $\Delta f_\rmn{H_2}$ are found by various other authors \citep{Wolfire10,Levrier12,Smith14,Gong18,Li18}, although, as stated before, the authors denote it as DGFs.

The results show that in some cases there is a significant amount of (diffuse) H$_2$ gas in low-$I_\rmn{CO}$ regions, which can be problematic in actual observations: Even when neglecting complications of intrinsically CO-dark gas in regions with $I_\rmn{CO}$ $>$ 0.1 K km s$^{-1}$, i.e. assuming that the amount of H$_2$ in those regions ($M_\rmn{H_2}^\rmn{CO}$) can be determined accurately from CO, the observations would miss a significant amount of H$_2$ in the clouds due to the sensitivity limit.

\subsubsection{The $X_\rmn{CO}$-factor}
\label{sec:XCO}

\begin{figure*}
 \includegraphics[width=\textwidth]{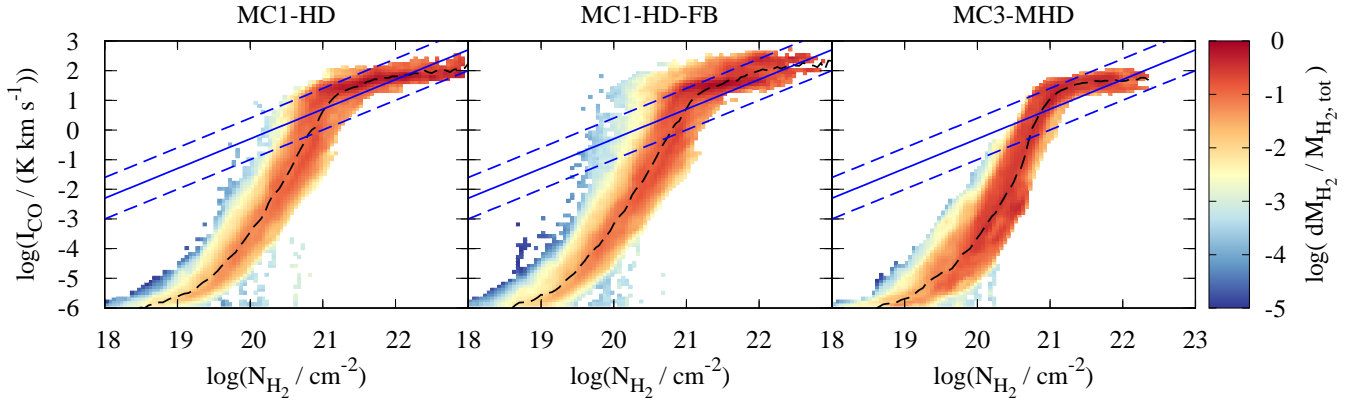}
 \caption{H$_2$-mass-weighted $N_\rmn{H_2}$ - $I_\rmn{CO}$-phase diagram of the clouds MC1-HD (left), MC1-HD-FB (middle), and MC3-MHD (right) along the $x$-direction at $t_\rmn{evol}$ = 3 Myr. Overall, there are only marginal differences between the different runs irrespective of the absence or presence of either magnetic fields or radiative stellar feedback. The dashed black line shows the mean of the distribution, the blue lines are obtained when using a canonical value of the $X_\rmn{CO}$-factor of \mbox{2 $\times$ 10$^{20}$ cm$^{-2}$ K$^{-1}$ km$^{-1}$} (solid) and 5 times higher and lower values (dashed).}
 \label{fig:XCO}
\end{figure*}
\begin{figure*}
 \includegraphics[width=\textwidth]{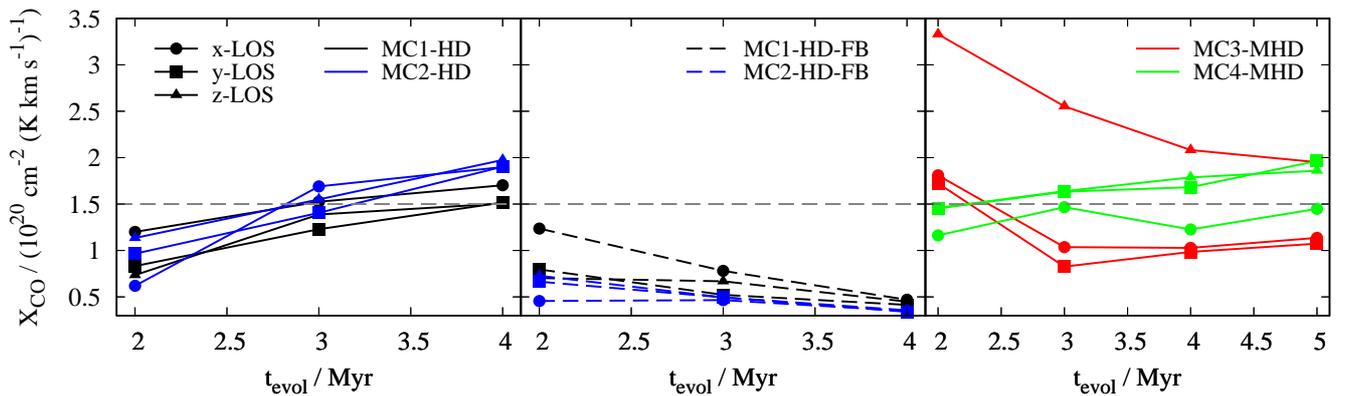}
 \caption{Time evolution of the $X_\rmn{CO}$-factor of the regions with $I_\rmn{CO} >$ 0.1 K km s$^{-1}$ for the various runs. There is a significant scatter of a factor of a few around a value of \mbox{1.5 $\times$ 10$^{20}$ cm$^{-2}$ (K km s$^{-1}$)$^{-1}$} (grey dashed line), with the runs including feedback (middle panel) having systematically lower values.}
 \label{fig:XCO_tot}
\end{figure*}

In order to obtain $M_\rmn{H_2}$ of MCs from CO(1-0) observations, typically a fixed conversion factor, the so-called $X_\rmn{CO}$-factor is used, such that 
\begin{equation}
 N_\rmn{H_2} = X_\rmn{CO} \times I_\rmn{CO} \, .
\end{equation}
The canonical value of $X_\rmn{CO}$ in the MilkyWay is assumed to be about \mbox{2 $\times$ 10$^{20}$ cm$^{-2}$ (K km s$^{-1}$)$^{-1}$} \citep[][but see also the review by \citealt{Bolatto13}]{Dame93,Strong96}. However, as the DGF is varying strongly among the MCs (Section~\ref{sec:dgf_total}), this raises the question to what extent the $X_\rmn{CO}$-factor is affected as well.

For this reason, we first investigate the relation between the H$_2$ column density, $N_\rmn{H_2}$, and the simulated CO(1-0) intensity for three representative clouds MC1-HD, MC1-HD-FB, and MC3-MHD along the $x$-direction at $t_\rmn{evol}$ = 3 Myr  (Fig.~\ref{fig:XCO}). Overall, we find only little variation in the general functional shape when including either magnetic fields or stellar feedback and when considering different directions or times (with the latter two not shown here), although for the runs with feedback the distribution becomes somewhat broader. For all runs we find a strong increase in $I_\rmn{CO}$ above \mbox{$N_\rmn{H_2}$ $\sim$ a few $\times$ 10$^{19}$ cm$^{-2}$} \citep[see also][for observational examples]{Federman80,Liszt98,Sheffer08}. Above \mbox{$N_\rmn{H_2}$ $\simeq$ 10$^{21}$ cm$^{-2}$}, the maximum CO intensity saturates around \mbox{$I_\rmn{CO}$ $\sim$ 100 K km s$^{-1}$}, where also most of the H$_2$ mass sits. This saturation is also seen in other theoretical and observational works \citep[e.g.][]{Pineda08,Ripple13,Smith14,Gong18} and can be attributed to the fact that CO(1-0) becomes optically thick already around an $A_\rmn{V, 2D}$ of $\sim$~1 \citep{Seifried17}.

For comparison, we also show the relation obtained when using a fixed $X_\rmn{CO}$-factor to convert $I_\rmn{CO}$ to $N_\rmn{H_2}$ (blue lines). Overall, there is a scatter of up to several orders of magnitude around this relation, which is in agreement with our previous results \citep[][but see also e.g. \citealt{Smith14,Gong18}]{Seifried17}. Furthermore, the distribution of $N_\rmn{H_2}$ and $I_\rmn{CO}$ shows an (almost) linear relation -- required for X$_\rmn{CO}$ to be applicable -- only for a small range of column densities around \mbox{$N_\rmn{H_2}$ $\simeq$ 10$^{21}$ cm$^{-2}$}. For runs with stellar feedback this range is somewhat more extended to higher column densities, as here a significant amount of CO gets destroyed (Section~\ref{sec:dgf_total}) and thus CO(1-0) becomes optically thick, i.e. the $N_\rmn{H_2}$-$I_\rmn{CO}$ relation becomes flat, only at higher $N_\rmn{H_2}$. Overall, however, our findings agree with previous results that on sub-pc scales, i.e. for individual pixels, the $X_\rmn{CO}$-factor is not applicable \citep[e.g.][]{Glover11,Shetty11a,Shetty11b,Bolatto13,Finn19}.

In Fig.~\ref{fig:XCO_tot} we show the $X_\rmn{CO}$-factor obtained by integrating both the column density and the CO(1-0) intensity over the observable regions, i.e. where $I_\rmn{CO} >$ 0.1 K km s$^{-1}$. There are clear differences between the different MCs recognisable, which directly translates into uncertainties in the inferred H$_2$ cloud mass. Most prominently, as already found in \citet{Glover16} and \citet{Seifried17}, the $X_\rmn{CO}$-factor of the hydrodynamical runs without feedback (left panel) increases over time with typical values around \mbox{0.5 -- 2 $\times$ 10$^{20}$ cm$^{-2}$ (K km s$^{-1}$)$^{-1}$}.

In contrast to that, for the runs including magnetic fields (right panel), $X_\rmn{CO}$ partly decreases over time with typical values from \mbox{0.8 -- 4 $\times$ 10$^{20}$ cm$^{-2}$ (K km s$^{-1}$)$^{-1}$} \citep[see also][]{Richings16a,Richings16b}. However, towards later stages the values appear to converge around \mbox{1 -- 2 $\times$ 10$^{20}$ cm$^{-2}$ (K km s$^{-1}$)$^{-1}$}. We find that more diffuse clouds (here the MHD clouds, see Fig.~\ref{fig:AV_pdf}) tend to have somewhat higher $X_\rmn{CO}$-factors than more compact clouds (here the HD clouds). This is in good agreement with observational findings for the Perseus, Taurus, and Orion molecular cloud \citep{Pineda08,Pineda10,Ackermann12,Lee14} showing lower $X_\rmn{CO}$-factors for denser and more compact sub-regions \citep[see also][for similar numerical results]{Glover11,Szucs16,Seifried17}.

The values of $X_\rmn{CO}$ of the runs with feedback (middle panel of Fig.~\ref{fig:XCO_tot}) are typically somewhat lower \mbox{($\lesssim$ 1 $\times$ 10$^{20}$ cm$^{-2}$ (K km s$^{-1}$)$^{-1}$)} than those of the runs without feedback. We attribute this to the fact that in these runs CO(1-0) becomes optically thick later (middle panel of Fig.~\ref{fig:XCO}) and thus the amount of CO intensity for a given amount of H$_2$ is higher than that for a run without feedback.

Overall, our average value for $X_\rmn{CO}$ is around \mbox{1.5 $\times$ 10$^{20}$ cm$^{-2}$ (K km s$^{-1}$)$^{-1}$} in good agreement with other theoretical works \citep[e.g.][]{Glover11,Smith14,Duarte15,Glover16,Richings16a,Richings16b,Szucs16,Gong18,Li18}, although these works tend to have spatial resolutions coarser than the required limit of 0.1 pc \citep{Seifried17,Joshi19} or simplified descriptions of the chemical evolution. However, our results also show that the actual $X_\rmn{CO}$-factor can vary by up to a factor of $\sim$ 4 in either direction for different MCs. This in turn implies the \textit{same} uncertainty of a factor of $\sim$ 4 for the inferred H$_2$ cloud masses. As e.g. the virial parameter scales with $M_\rmn{H_2}^2$, this can result in uncertainties of one order of magnitude for inferred quantities. We note that the partly significant cloud-to-cloud variations of $X_\rmn{CO}$ reported here are in good agreement with variations reported over decades in observations of galactic and extra-galactic MCs \citep[e.g.][but see also the review by \citealt{Bolatto13}]{Blitz80,Scoville87,Dame93,Strong96,Melchior00,Lombardi06,Nieten06,Leroy11,Smith12,Ripple13} and the theoretical works noted before. This indicates that, besides being not applicable on sub-pc scales, the $X_\rmn{CO}$-factor might have its strength when being applied for an ensemble of MCs rather than individual MCs \citep[e.g.][]{Kennicutt12}.

\section{Towards a new approach to determine the H$_2$ content of molecular clouds}
\label{sec:newapproach}

\begin{figure}
 \includegraphics[width=\linewidth]{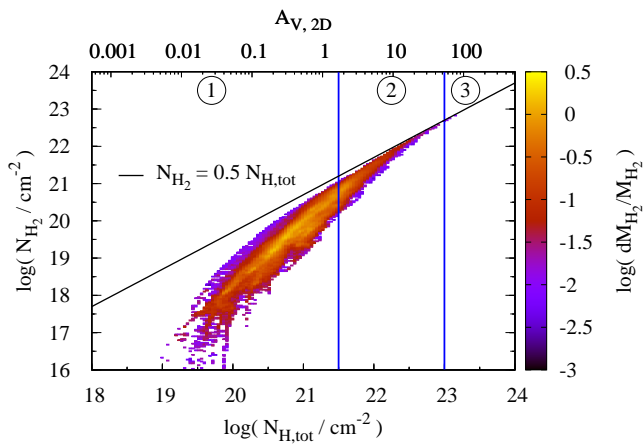}
 \caption{Dependence of $N_\rmn{H_2}$ on the total hydrogen column density $N_\rmn{H, tot}$ (lower $x$-axis) and the corresponding $A_\rmn{V, 2D}$ value (upper $x$-axis) for run MC1-HD projected along the $x$-direction at $t_\rmn{evol}$ = 3 Myr. The blue vertical lines and encircled numbers indicate the three regimes used for the new approach. Above $N_\rmn{H}$ $\simeq$ 10$^{21.5}$~cm$^{-2}$ there is a good correlation with $N_\rmn{H_2}$ used for regime (2) and (3) of the new approach (Eqs.~\ref{eq:fitmid} and~\ref{eq:fithigh})}
 \label{fig:NH2}
\end{figure}

\begin{figure*}
 \includegraphics[width=0.48\linewidth]{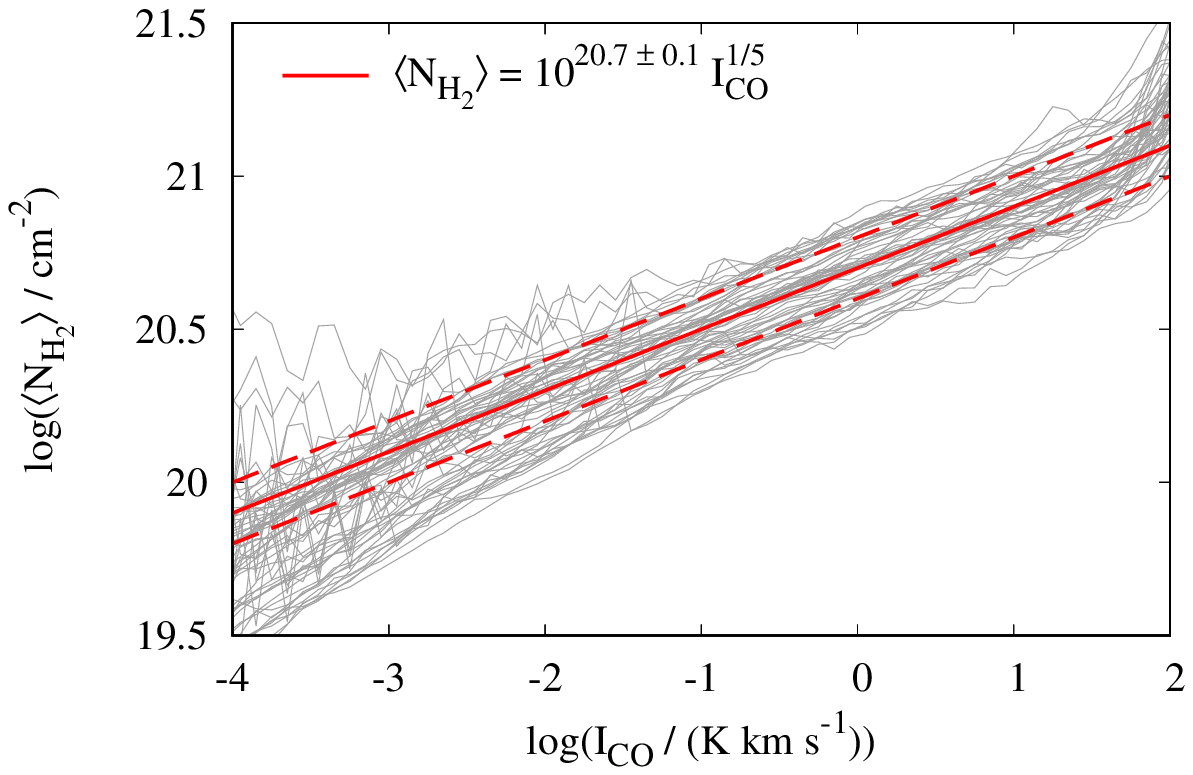}
 \includegraphics[width=0.48\linewidth]{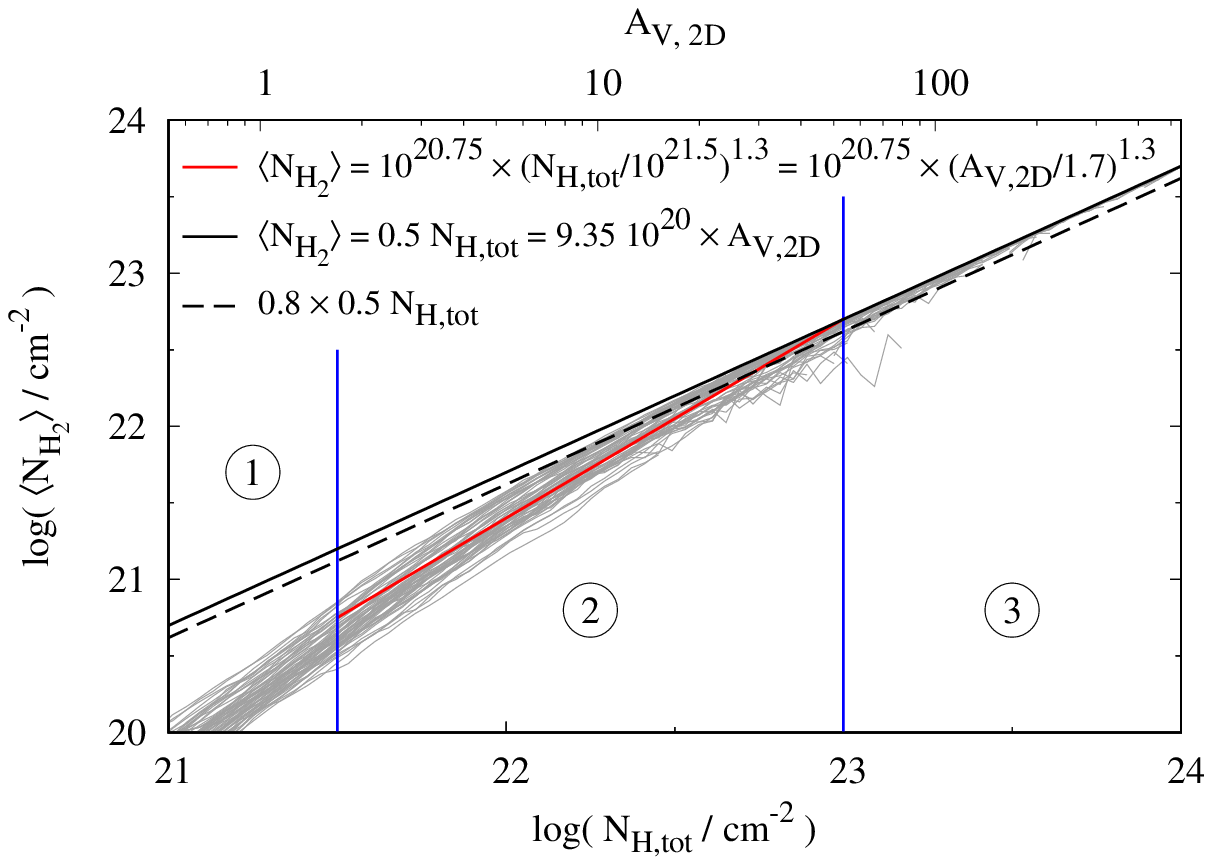}
 \caption{Mean value of $N_\rmn{H_2}$ as a function of $I_\rmn{CO}$ (left) and of $N_\rmn{H, tot}$ (right) for all runs and all three directions at $t_\rmn{evol}$ = 2, 3, 4, and 5 Myr (grey lines).
 Left: $N_\rmn{H_2}$ can be well approximated by the CO intensity from 10$^{-2}$ to 10 K km s$^{-2}$ with a typical uncertainty of 0.1 dex (red lines). This is used for \mbox{$N_\rmn{H, tot} < 10^{21.5}$ cm$^{-2}$}, i.e. regime (1) of the new approach (Eq.~\ref{eq:fitlow}). Right: The blue lines and encircled numbers indicate the regimes of the new approach. At higher values, $N_\rmn{H_2}$ can be inferred from extinction measurements. For $10^{21.5}$ cm$^{-2}$ $\leq  N_\rmn{H, tot} < 10^{23}$ cm$^{-2}$, i.e. regime (2) of the new approach (Eq.~\ref{eq:fitmid}), the gas is not yet fully molecular (red line). For $N_\rmn{H, tot} > 10^{23}$ cm$^{-2}$, i.e. regime (3) (Eq.~\ref{eq:fithigh}), the gas can be assumed to be fully molecular (black line).}
 \label{fig:fits}
\end{figure*}

In the literature several ways are commonly used to estimate the (molecular) mass of MCs. As described before, the often used $X_\rmn{CO}$-factor is subject to significant cloud-to-cloud variations, which in turn impose a factor-of-a-few uncertainty for the H$_2$ mass in any MC (neglecting the CO-faint H$_2$ gas discussed in Section~\ref{sec:limit}). Furthermore, at low column densities ($N_\rmn{H_2}$ $\lesssim$ 10$^{21}$ cm$^{-2}$), the strong correlation between CO and H$_2$ (see Fig.~\ref{fig:XCO}) has been used to estimate the H$_2$ content of MCs \citep[e.g.][]{Federman80,Liszt98,Sheffer08}. This approach, however, breaks down at high column densities, where CO becomes optically thick. In this latter regime, the usage of extinction maps in combination with Eq.~\ref{eq:AV2D} can provide the hydrogen mass of MCs. However, this mass describes the total hydrogen mass (atomic hydrogen \textit{and} H$_2$) and should thus not be mixed up with the actual molecular H$_2$ mass, as the gas might not necessarily be fully molecular (Fig.~\ref{fig:NH2}). Hence, it is questionable to which extent $A_\rmn{V, 2D}$ can be used to assess the molecular content/chemical state of an MC; a question, which we will investigate further below.

In the following we describe a new approach, which tries to reduce the shortcomings of the aforementioned approaches and is based on three column density regimes (or alternatively $A_\rmn{V, 2D}$ regimes, Eq.~\ref{eq:AV2D}). We define these regimes in Fig.~\ref{fig:NH2}, which shows the relation between $N_\rmn{H_2}$ and $N_\rmn{H, tot}$ for MC1 at $t$ = 3 Myr. Above \mbox{$N_\rmn{H, tot}$ $\simeq$ 10$^{21.5}$ cm$^{-2}$}, $N_\rmn{H_2}$ shows a good correlation with $N_\rmn{H, tot}$, whereas below there is a scatter of half an order of magnitude and more. For this reason, for regime (1) we rely on CO(1-0) observations and for the regimes (2) and (3) we rely on extinction measurements. In the following, we first describe in detail the principles of the approach, before we provide the actual numbers and interpret the results.
\begin{enumerate}
 \item[(1)] $N_\rmn{H, tot} < 10^{21.5}$ cm$^{-2}$:\\
As shown in Fig.~\ref{fig:XCO}, for values between $I_\rmn{CO}$ $\simeq$ 10$^{-4}$ -- \mbox{10 K km s$^{-1}$}, there is a good correlation between $N_\rmn{H_2}$ and $I_\rmn{CO}$, which allows us to express $N_\rmn{H_2}$ as a function of $I_\rmn{CO}$. For this purpose, in the left panel of Fig.~\ref{fig:fits} we show the mean H$_2$ column density, $\left\langle N_\rmn{H_2} \right\rangle$, as a function of $I_\rmn{CO}$ for all runs at all times and all three LOS considered. For a given CO intensity, the values of $\left\langle N_\rmn{H_2} \right\rangle$ vary by about 0.1 dex, i.e. 25\% in either direction. Given this good qualitative agreement among the different runs, we approximate $\left\langle N_\rmn{H_2} \right\rangle$ by a powerlaw (see Eq.~\ref{eq:fitlow} below) where we focus on matching the curves in the range 10$^{-2}$ K km s$^{-1}$ $<$ $I_\rmn{CO}$ $<$ 10 K km s$^{-1}$.
\item[(2)] $10^{21.5}$ cm$^{-2}$ $\leq  N_\rmn{H, tot} < 10^{23}$ cm$^{-2}$:\\
As the correlation between $I_\rmn{CO}$ and $N_\rmn{H_2}$ breaks down at \mbox{$N_\rmn{H_2} \gtrsim 10^{21}$~cm$^{-2}$} ($I_\rmn{CO}$ $\gtrsim$ 10 K km s$^{-1}$, Fig.~\ref{fig:XCO}), we next consider the relation between $N_\rmn{H_2}$ and $N_\rmn{H, tot}$, which is accessible through dust extinction measurements. This is shown exemplarily in Fig.~\ref{fig:NH2} for MC1-HD along the $x$-direction at $t_\rmn{evol}$ = 3 Myr. We find that for $N_\rmn{H, tot}$ $\geq$ 10$^{21.5}$~cm$^{-2}$ there is a good correlation between both quantities. For this reason, in the right panel of Fig.~\ref{fig:fits} we plot $\left\langle N_\rmn{H_2} \right\rangle$ as a function of $N_\rmn{H, tot}$ (and $A_\rmn{V, 2D}$) for all runs, times and three LOS considered. In the range $10^{21.5}$~cm$^{-2}$ $\leq$  $N_\rmn{H, tot}$ $<$ 10$^{23}$~cm$^{-2}$, $\left\langle N_\rmn{H_2} \right\rangle$ shows only moderate deviations between the individual runs of $\pm$ 0.2 dex, i.e. a factor of 1.6 in either direction and an almost linear trend in double-logarithmic representation. For this reason, we apply a powerlaw in this range (see Eq.~\ref{eq:fitmid} below), which smoothly connects to the range where $N_\rmn{H_2}$ and $I_\rmn{CO}$ are still correlated (regime (1)).
\item[(3)] $N_\rmn{H, tot} \geq 10^{23}$ cm$^{-2}$:\\
For $N_\rmn{H, tot} \geq 10^{23}$ cm$^{-2}$ ($A_\rmn{V, 2D} \geq 53.5$), the gas becomes (almost) fully molecular within typical deviations of $<$ 20\% (black dashed curve in the right panel of Fig.~\ref{fig:fits}). We thus approximate $N_\rmn{H_2}$ by 0.5 $\times$ $N_\rmn{H, tot}$ (see Eq.~\ref{eq:fithigh} below).
\end{enumerate}

\subsection{The final approach}

Taking all results together, we obtain a new way to calculate $N_\rmn{H_2}$ which is based solely on CO(1-0) and visual extinction observations:
\begin{enumerate}
 \item[(1)] For $\frac{N_\rmn{H, tot}}{\textrm{1 cm$^{-2}$}} < 10^{21.5} \textrm{ or } A_\rmn{V, 2D} < 1.7$:
 \begin{equation}
  N_\rmn{H_2} = 10^{20.7} \textrm{cm$^{-2}$} \times \left(\frac{I_\rmn{CO}}{\textrm{K km s$^{-1}$}}\right)^{1/5} \label{eq:fitlow}
 \end{equation}
 \item[(2)] For $10^{21.5} \leq  \frac{N_\rmn{H, tot}}{\textrm{1 cm$^{-2}$}} < 10^{23} \textrm{ or } 1.7 \leq A_\rmn{V, 2D} < 53.5$:
 \begin{equation}
  N_\rmn{H_2} = 10^{20.75} \textrm{cm$^{-2}$} \times \left(\frac{N_\rmn{H, tot}}{10^{21.5} \textrm{cm$^{-2}$}}\right)^{1.3} = 10^{20.75} \textrm{cm$^{-2}$} \times \left(\frac{A_\rmn{V, 2D}}{1.7}\right)^{1.3} \label{eq:fitmid}
 \end{equation}
 \item[(3)] For $\frac{N_\rmn{H, tot}}{\textrm{1 cm$^{-2}$}} \geq 10^{23} \textrm{ or } A_\rmn{V, 2D} \geq 53.5$:
 \begin{equation}
  N_\rmn{H_2} = 0.5 \times N_\rmn{H, tot} = 0.935 \times 10^{21} \textrm{cm$^{-2}$} \times A_\rmn{V, 2D} \label{eq:fithigh}
 \end{equation}
\end{enumerate}

Our approach is applicable for a wide range column densities. In the intermediate- to high-$A_\rmn{V, 2D}$ range, it relies on the availability of extinction measurements in order to minimize the uncertainties introduced by the variations in $X_\rmn{CO}$. At $10^{21.5}$~cm$^{-2}$ $\leq$ $N_\rmn{H, tot}$ $<$ 10$^{23}$ cm$^{-2}$, $N_\rmn{H_2}$ we introduce a novel super-linear relation between $A_\rmn{V, 2D}$ and $N_\rmn{H_2}$ as the gas becomes increasingly molecular. Our approach also covers the low-$A_\rmn{V, 2D}$ range, which is accessible via high-sensitivity, long-exposure CO observations. Here, the weak dependence of $N_\rmn{H_2}$ on $I_\rmn{CO}$ in Eq.~\ref{eq:fitlow} -- or inversely the strong dependence of $I_\rmn{CO}$ on $N_\rmn{H_2}$ -- can be understood as a consequence of CO being underabundant (with respect to H$_2$) at low column densities and then rapidly builds up to ``catch up'' with H$_2$ towards higher column densities. This is also indicated in the left panel of Fig.~\ref{fig:phasediag} \citep[see also][]{Gong18}, where we indeed find a steep rise of $f_\rmn{CO}$ above $f_\rmn{H_2}$ $\simeq$ 0.5. Finally, we note that using somewhat different fitting ranges in Eqs.~\ref{eq:fitlow}-\ref{eq:fithigh} results in changes for $N_\rmn{H_2}$ of a few percent only.

\subsection{Comparison of the new approach and previous approaches}
\label{sec:comparison}

In order to test the applicability of the suggested approach, we apply it to our simulations to determine the total H$_2$ mass in the clouds and compare it to the classical approach of a constant $X_\rmn{CO}$-factor. In addition, we compare it to the approach of \citet{Glover11}, which suggest that the $X_\rmn{CO}$-factor varies with the visual extinction averaged over the entire cloud, $\left\langle A_\rmn{V, 2D}\right\rangle$, as
\begin{equation}
 X_\rmn{CO} = X_\rmn{CO, 0} \times \left(\frac{\left\langle A_\rmn{V, 2D}\right\rangle}{3.5}\right)^{-3.5} \, .
\end{equation}
Motivated by the findings of Fig.~\ref{fig:XCO_tot}, we use a value of \mbox{1.5 $\times$ 10$^{20}$ cm$^{-2}$ (K km s$^{-1}$)$^{-1}$} for $X_\rmn{CO, 0}$ as well as for the approach with a constant $X_\rmn{CO}$-factor. We constrain ourselves to the observable regions\footnote{We only focus on a particular threshold value as the total mass of H$_2$ depends only weakly on it (Fig.~\ref{fig:DGF_CO}).} where \mbox{$I_\rmn{CO}$ $>$ 0.1 K km s$^{-1}$}.

\begin{figure*}
 \includegraphics[width=\linewidth]{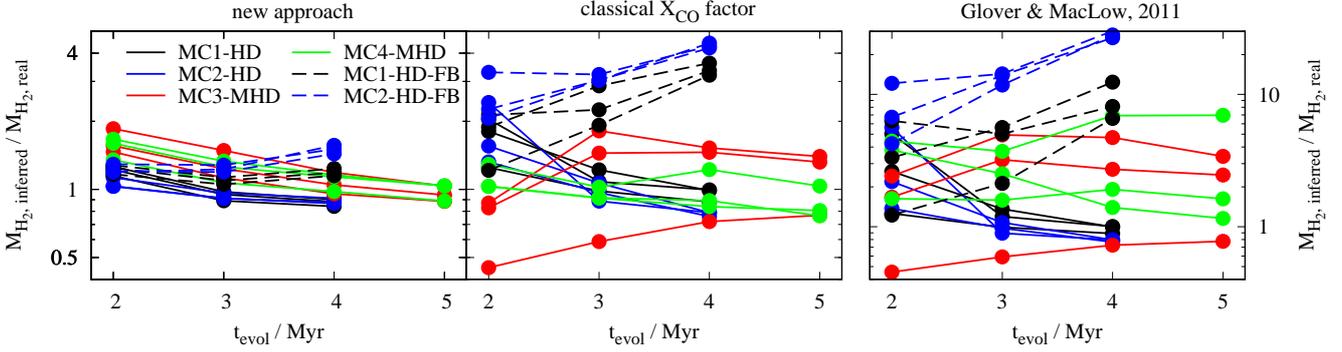} 
 \caption{Ratio of the estimated H$_2$ mass to the actual H$_2$ mass for the different MCs and directions as a function of time. The new approach suggested in this work (Eqs.~\ref{eq:fitlow}--\ref{eq:fithigh}) predicts H$_2$ masses ranging from 80\% to 180\% of the actual masses (left), whereas for the classical approach via an $X_\rmn{CO}$-factor of \mbox{1.5 $\times$ 10$^{20}$ cm$^{-2}$ (K km s$^{-1}$)$^{-1}$} the masses are uncertain within a factor of 3 -- 4 in either direction (middle). The approach suggested by \citet{Glover11} (right) results in even larger deviations (note the different $y$-axis scaling here).}
 \label{fig:h2masses}
\end{figure*}
In Fig.~\ref{fig:h2masses} we compare the inferred H$_2$ masses to the actual H$_2$ masses for the various MCs, directions and approaches. Overall, we find that our new approach (left panel) matches the actual H$_2$ mass best: the estimated H$_2$ masses range from 80\% to 180\% of the actual masses, i.e. deviations of a factor of 1.8 at most. This is about 2 times better than with the classical approach via a constant $X_\rmn{CO}$-factor, which shows deviations of a factor of 3 -- 4 in either direction (middle panel). The approach of \citet{Glover11} results in even larger deviations of up to one order of magnitude, in particular for the runs with feedback (right panel). Similar deviations, in particular the rather poor match for the approach of \citet{Glover11}, were also found by \citet{Szucs16}. We speculate that this is due to the fact that \citet{Glover11} use turbulent box simulations, which are difficult to compare to our simulations and might also be problematic when considering the convergence of the chemical abundances \citep{Joshi19}. In addition, the authors only approximate the CO(1-0) intensity by a simplified radiative transfer scheme, which could also contribute to the observed differences.

We emphasize that our new approach works equally well for all three different situations considered in this paper, i.e. clouds with and without magnetic fields and stellar feedback. There is, however, a slight tendency for the runs without feedback to overestimate the H$_2$ mass at early times and to underestimate it at late times (vice versa for the runs with feedback), which we investigate further below.

\begin{figure*}
 \includegraphics[width=\linewidth]{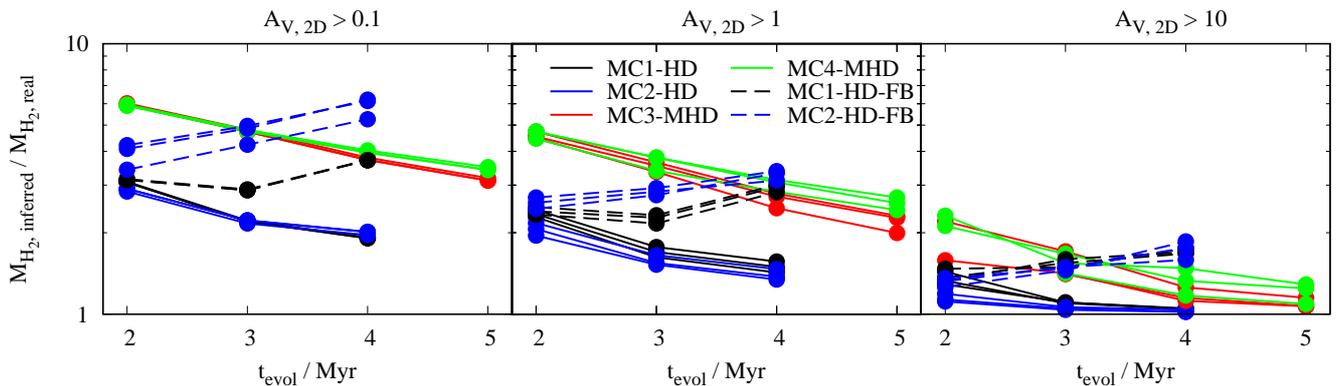} 
 \caption{Ratio of the (total) hydrogen mass calculated via Eq.~\ref{eq:AV2D} to the actual H$_2$ mass for the different MCs and directions as a function of time. Only pixels above a threshold of $A_\rmn{V, 2D}$ = 0.1, 1, and 10 (from left to right) are considered. The ratio approaches unity as the threshold increases.}
 \label{fig:hmasses}
\end{figure*}
Next, we compare the (total) hydrogen masses of the various MCs obtained via Eq.~\ref{eq:AV2D} to their H$_2$ masses, only taking into account pixels above a threshold of $A_\rmn{V, 2D}$ = 0.1, 1, and 10, respectively (Fig.~\ref{fig:hmasses}). As expected, the ratio approaches unity with increasing threshold since the gas becomes increasingly more molecular (see Fig.~\ref{fig:NH2}). However, for typical observational sensitivity thresholds around $A_\rmn{V, 2D}$ = 0.1 -- 1, the total mass is up to a factor of $\sim$~7 higher than the H$_2$ mass and should therefore not be used to estimate the molecular gas content. Even when considering only the densest parts of the clouds ($A_\rmn{V, 2D}$ $>$ 10), $M_\rmn{H_2}$ would be overestimated by a few 10\% up to a factor of 2. This is a consequence of the fact that the gas becomes (on average) fully molecular only above an $A_\rmn{V, 2D}$ of a few times 10 (our regime (3), see Fig.~\ref{fig:NH2}), which our approach accounts for by introducing the non-linear relation between N$_\rmn{H_2}$ and $A_\rmn{V, 2D}$ in regime (2) (Eq.~\ref{eq:fitmid}).

\subsection{The accuracy of the approach}
\label{sec:accuracy}

\begin{figure}
 \includegraphics[width=\linewidth]{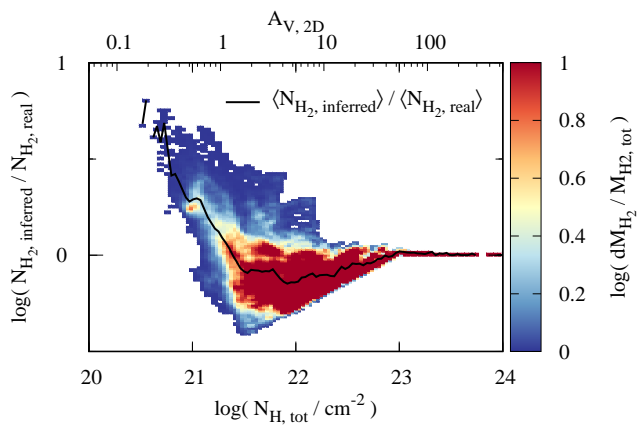} 
 \caption{Ratio of the estimated H$_2$ to the actual H$_2$ column density for each pixel of run MC1-HD projected along the $z$-direction at $t_\rmn{evol}$ = 3 Myr.}
 \label{fig:h2pixels}
\end{figure}

There are two main sources of uncertainty of the suggested approach, first the simulations themselves (Section~\ref{sec:caveats}), and second the combination of different runs and times to obtain the final fits. In order to explore the effect of the latter, we calculate the ratio of the inferred and actual H$_2$ column density in each pixel of the projected 2D-maps (again considering only pixels with $I_\rmn{CO} >$ 0.1 K km s$^{-1}$) as a function of  $N_\rmn{H, tot}$ ($A_\rmn{V, 2D}$) and plot the resulting H$_2$-mass-weighted phase diagram of MC1-HD projected along the $z$-direction at $t_\rmn{evol}$ =  3 Myr (Fig.~\ref{fig:h2pixels}). For each $N_\rmn{H, tot}$-bin, we also calculate the mean of the inferred and actual H$_2$ column densities, $\left \langle N_\rmn{H_2, inferred} \right \rangle$ and $\left \langle N_\rmn{H_2, real} \right \rangle$, respectively, and plot their ratio.

We find that our approach matches the actual H$_2$ column density even on an individual pixel basis relatively well, which also holds for the other MCs, times and directions. In regime (1) and (2), the typical deviations are of the order of at most \mbox{$\pm$ 0.5 dex}, i.e. lower than a factor of $\sim$ 3. Towards the highest $N_\rmn{H, tot}$ (regime (3)), where most of the H$_2$ mass resides, the ratio is close to 1. These deviations are in rough agreement with those of the underlying correlations (Figs.~\ref{fig:XCO} and~\ref{fig:NH2}) and the deviations of their mean values from the applied fit (Fig.~\ref{fig:fits}).

Hence, overall the uncertainties in our approach appear to be dominated by variations in the intermediate column density range, i.e. regime (2) of the approach (Eq.~\ref{eq:fitmid}), where a significant  amount of the gas resides (Fig.~\ref{fig:h2pixels}). The typical deviations of the fitted from the real $N_\rmn{H_2}$ in this column density regime (see also right panel of Fig.~\ref{fig:fits}) also roughly match the overall accuracy of our approach of a factor of $\lesssim$ 1.8 for the total H$_2$ mass. Also the time trends seen in the left panel of Fig.~\ref{fig:h2masses} are mainly caused by (time) variations in this regime. However, as we aim at providing a method which is valid to estimate H$_2$ for \textit{various} evolutionary stages, this uncertainty cannot be further reduced.  We note, however, that the match for the $X_\rmn{CO}$-factor (not shown, but see Fig.~\ref{fig:XCO}), would be significantly worse with typical deviations of a factor of 10 (or even more) in either direction.

\subsection{Caveats}
\label{sec:caveats}

The fits to determine the H$_2$ mass (Eqs.~\ref{eq:fitlow}--\ref{eq:fithigh}) are strictly seen only valid for an ISRF and a cosmic ray ionisation rate (CRIR) corresponding to solar neighborhood conditions, i.e. $G_0$ = 1.7 and CRIR = 1.3 $\times$ 10$^{-17}$ s$^{-1}$. However, \citet{Clark15} show that the $X_\rmn{CO}$-factor, and thus $I_\rmn{CO}$, for MCs with masses around 10$^4$~M$_{\sun}$ is barely affected when varying the ISRF and CRIR over two orders of magnitude. Also for more massive MCs ($\sim$ 10$^5$~M$_{\sun}$), they find only a relatively weak dependence on these quantities. In the densest and thus UV-shielded regions, where cosmic rays are expected to have a larger impact on the CO-H$_2$ ratio \citep{Bisbas15,Bisbas17}, our approach relies on the extinction measurement, thus being less sensitive to the CRIR. Furthermore, \citet{Glover11} and \citet{Shetty11a} show that also the influence of a moderately varying metallicity on $X_\rmn{CO}$ is rather limited \citep[see also][]{Szucs16}.

Uncertainties in the calculation of the UV shielding by surrounding gas can also affect the calculation of molecular abundances. The assumed converstion factor of \mbox{5.348$\times$10$^{-22}$ cm$^2$ mag} (Eq.~\ref{eq:AV2D}) shows cloud-to-cloud variations of several 10\% \citep[see Fig.~2 of ][]{Bohlin78}, which would affect the dissociation of H$_2$ and CO (i.e. regime (1)) but also directly the conversion of observed extinction into $N_\rmn{H, tot}$ (regimes (2) and (3)). However, as shown by \citet{Glover11} and \citet{Szucs16}, the H$_2$ content depends only weakly on the extinction and varies only by a few 10\% even when varying the ISRF by a factor of 10. Furthermore, the error introduced by the discretisation within the {\sc TreeCol/OpticalDepth} algorithm into 48 directions (Eq.~\ref{eq:AV}) is typically only of the order of 1 -- 10\% \citep{Clark12}. Taken together, we thus speculate that the formulae given in Eqs.~\ref{eq:fitlow}--\ref{eq:fithigh} might still be applicable even under somewhat different environmental conditions than in the solar neighborhood.

We also note that the used chemical network \citep{Nelson97,Glover07,Glover10} is simplified to allow for an efficient application in 3D, MHD simulations. However, comparison calculations with a more extended network based on \citet{Nelson99} and \citet[][see the appendix of \citealt{Mackey19} for details]{Glover10} show a reasonable agreement with the more simple network applied here. Furthermore, our results are also in agreement with the work of \citet{Levrier12}, which use a significantly more extended chemical network. The authors chemically post-process MHD simulations of \citet{Hennebelle08} with the Meudon Code \citep{LePetit06}, which constrains them to a plane-parallel geometry and equilibrium chemistry. Despite this quite different approach, their results agree well with ours concerning the importance of density enhancements along the LOS (Section~\ref{sec:2dmaps}) and the observational deficit $\Delta f_\rmn{H_2}$ (Section~\ref{sec:limit}), which makes us confident about our chosen network.

Finally, we note that so far for the radiative stellar feedback we have only considered a single energy band (all photons with energies above 13.6 eV). We are currently working on including additional energy bands, which would allow for an even more detailed description of the dissociation processes of H$_2$ and CO. Furthermore, for future simulations we plan to achieve an even higher spatial resolution in order to assure that in particular the CO content in our simulations is fully resolved \citep{Joshi19}.

\section{Conclusions}
\label{sec:conclusion}

We present high-resolution (0.1 pc) simulations of molecular cloud formation including a live chemical network for H$_2$ and CO as well as the necessary shielding processes. The simulations are part of the SILCC-Zoom project \citep{Seifried17} and include the galactic environment of the clouds, radiative stellar feedback and magnetic fields. We investigate six different simulations, 4 hydrodynamical runs, out of which 2 include star formation and ionisation feedback of young massive stars, and 2 magneto-hydrodynamical runs. In the simulations we can differentiate between the local visual extinction in each point, $A_\rmn{V, 3D}$, obtained directly from the 3D simulation data and the LOS-integrated visual extinction, $A_\rmn{V, 2D}$, as accessible in actual observations. In the following we list our main findings.
\begin{itemize}
\item The fraction of intrinsically CO-dark H$_2$ gas (DGF) varies from 40\% to 95\%, with higher values for magnetised MCs. We show that differences in the DGF can be attributed to the structure of the clouds: clouds with a high amount of CO-dark gas have less well-shielded gas. The DGF, however, does not correlate with the total H$_2$ mass.
\item CO-bright gas is typically found at hydrogen nuclei densities above \mbox{300 cm$^{-3}$}, temperatures below 50~K and local visual extinctions, $A_\rmn{V, 3D}$ $\gtrsim$ 1.5, where 50 -- 80\% or more of the total hydrogen and carbon atoms are in the form of H$_2$ and CO.
\item CO-dark gas extends into the more diffuse (\mbox{10 cm$^{-3}$} $\lesssim$ $n_\rmn{H,tot}$ $\lesssim$ \mbox{10$^3$ cm$^{-3}$}) and moderately cool gas (a few \mbox{$10$ K} $\lesssim$ T $\lesssim$ a few \mbox{100 K}). We speculate that this makes it difficult to probe the entire CO-dark gas with a single tracer. The typical $A_\rmn{V, 3D}$ of CO-dark gas ranges from 0.2 -- 0.3 to about 1 -- 1.5 independent of the presence or absence of either stellar feedback or magnetic fields. 
\item We demonstrate that with the LOS-integrated $A_\rmn{V, 2D}$, the conditions along the LOS cannot be determined properly. The actual distribution of the local visual extinction ($A_\rmn{V, 3D}$) along the LOS is broad and not in any way unique.
\item Related to that, we show that up to $A_\rmn{V, 2D}$ $\simeq$ 5, pixels can be CO-bright and CO-dark, i.e. the DGF is not well constrained by the observable visual extinction. This can be attributed to different density -- and thus $A_\rmn{V, 3D}$ -- distributions along the LOS: Pixels with a high DGF have a rather uniform density distribution with $A_\rmn{V, 3D}$~$\lesssim$~1 where CO is not formed. For CO-bright pixels, however, regions with strong density enhancements and locally well-shielded gas ($A_\rmn{V, 3D}$~$\gtrsim$~1.5) are present along the LOS.
\end{itemize}

In addition, we produced synthetic CO(1-0) observations of our simulated molecular clouds using RADMC-3D.
\begin{itemize}
\item We show that about 15 -- 65\% of the H$_2$ is in regions with \mbox{CO(1-0)} emission below an observational detection limit of \mbox{0.1 K km s$^{-1}$}, which amplifies the problem of \textit{intrinsically} CO-dark gas in regions with detectable emission. This fraction increases only slightly to 20 -- 75\% when a detection limit of \mbox{1 K km s$^{-1}$} is used.
\item We find a mean $X_\rmn{CO}$-factor of \mbox{$\sim$ 1.5 $\times$ 10$^{20}$ cm$^{-2}$ (K km s$^{-1}$)$^{-1}$} in our simulations with significant variations of a factor up $\sim$ 4 in good agreement with other observational and theoretical works. Hence, using $X_\rmn{CO}$ can result in significant errors in the estimated H$_2$ masses of individual clouds.
\end{itemize}

In order to overcome the long-standing problem to determine the H$_2$ content of MCs and to avoid the problem of CO-dark gas, we suggest a new approach to determine the H$_2$ content of MCs under solar neighborhood conditions. The approach relies on observations of the CO(1-0) line transition and the visual extinction. The formulae given in Eqs.~\ref{eq:fitlow}--\ref{eq:fithigh} present an approximation to the data obtained from all simulations considered here, which cover a variety of cloud conditions including and excluding both radiative stellar feedback and magnetic fields. Furthermore, the approach is applicable for a wide range of visual extinctions: from the low-extinction range ($A_\rmn{V, 2D}$ $\lesssim$ 1) covered by high-sensitivity, long-exposure CO observations, to the intermediate- and high-extinction range, where we introduce a novel a non-linear relation between $A_\rmn{V, 2D}$ and $N_\rmn{H_2}$.

The total H$_2$ cloud masses obtained with our new approach match the actual masses within a factor of at most 1.8 independent of whether feedback or magnetic fields are included or not. In contrast to that, the classical approach via a fixed $X_\rmn{CO}$-factor results in deviations of up to a factor of 4. Moreover, our approach also allows us to calculate the H$_2$ column density for individual pixels, i.e. on sub-pc scales, which is not possible with the $X_\rmn{CO}$-factor. Here, we find typical deviations from the real H$_2$ column density by less than a factor of 3, while the standard $X_\rmn{CO}$-factor results in deviations by an order of magnitude and more.

\section*{Acknowledgements}

The authors like to thank the anonymous referee for the comments which helped to significantly improve the paper. DS and SW acknowledge the support of the Bonn-Cologne Graduate School, which is funded through the German Excellence Initiative. DS, SH, SW and TGB also acknowledge funding by the Deutsche Forschungsgemeinschaft (DFG) via the Collaborative Research Center SFB 956 ``Conditions and Impact of Star Formation'' (subprojects C5 and C6). SW and TGB acknowledge support via the ERC starting grant No. 679852 "RADFEEDBACK". The FLASH code used in this work was partly developed by the Flash Center for Computational Science at the University of Chicago.
The authors acknowledge the Leibniz-Rechenzentrum Garching for providing computing time on SuperMUC via the project ``pr94du'' as well as the Gauss Centre for Supercomputing e.V. (www.gauss-centre.eu).




\bibliographystyle{mnras}
\bibliography{literature} 



\appendix
\section{Supplemental figures}

\subsection{Density and mass fractions}

\begin{figure}
 \includegraphics[width=\linewidth]{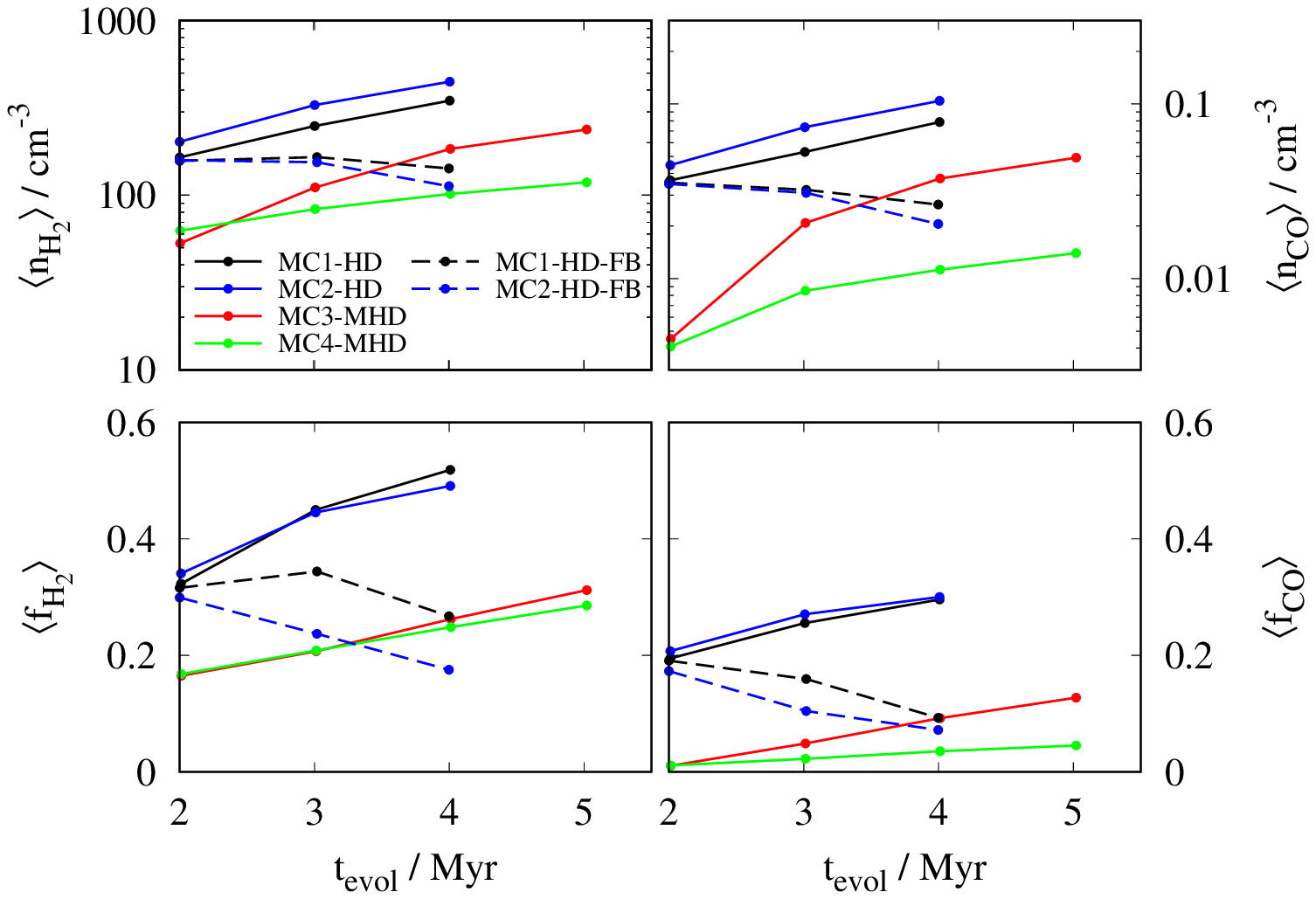}\\
 \caption{Time evolution of the mean H$_2$ and CO density in gas with total particle densities above 100 cm$^{-3}$ (top row) and the global mass fractions of H$_2$ and CO in the zoom-in regions (bottom row).}
 \label{fig:density}
\end{figure}

In the top row of Fig.~\ref{fig:density} we show the mean H$_2$ and CO densities in the dense gas, i.e only taking into account gas which has a density above \mbox{3.84 $\times$ 10$^{-22}$ g cm$^{-3}$} corresponding to a particle density of \mbox{$n$ = 100 cm$^{-3}$} for \mbox{$\mu$ = 2.3.} The MHD clouds show somewhat lower H$_2$ densities than the HD clouds, confirming the results of Section~\ref{sec:AV3D} that the MHD clouds are somewhat more diffuse. Similar holds true for the CO densities, which, due to the assumed fractional abundance of carbon atoms of 1.4 $\times$ 10$^{-4}$, are roughly a factor of 10$^{-4}$ lower.

In the bottom row we show the mass fractions of H$_2$ and CO in the \textit{entire} zoom-in regions. The mass fractions relate to the DGF (Eq.~\ref{eq:dgf}) as
\begin{equation}
 \textrm{DGF} = 1 - \frac{\left\langle f_\rmn{CO} \right\rangle}{\left\langle f_\rmn{H_2} \right\rangle} \, .
\end{equation}
As already indicated by the values of DGF $>$ 0 (Fig.~\ref{fig:dgf_total}) the mass fractions of CO are smaller than that of H$_2$. Furthermore, due to the more diffuse structure of the MHD clouds, in general both mass fractions are lower than that of the clouds without magnetic fields.

\subsection{Density-temperature phase diagrams}

\begin{figure*}
 \includegraphics[width=\textwidth]{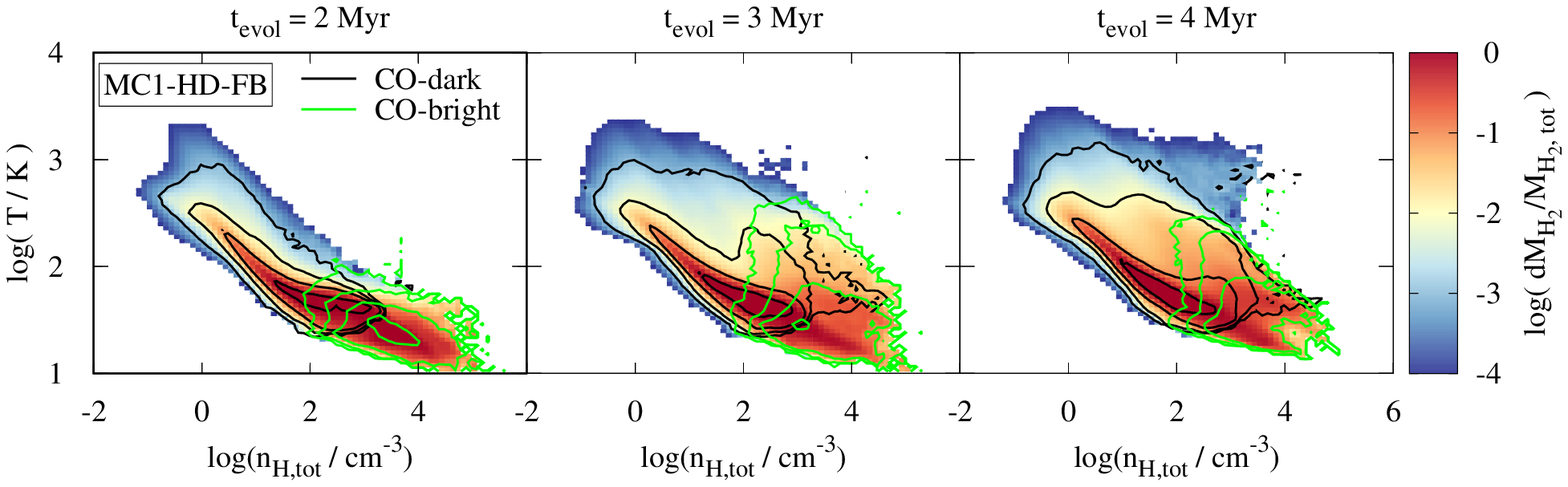}\\
 \includegraphics[width=\textwidth]{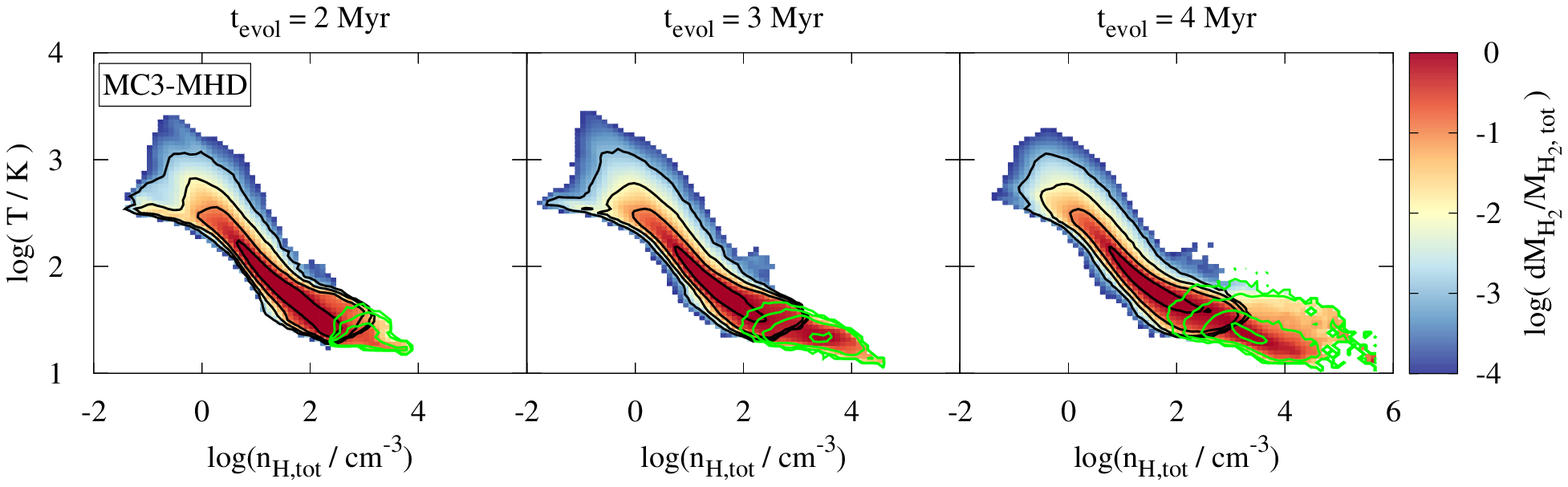}\\
 \caption{As in Fig.~\ref{fig:temp_dens_diag} but now for run MC1-HD-FB (top) and MC3-MHD (bottom). For the runs MC2-HD, MC2-HD-FB and MC4-MHD (not shown) we find very similar results. Feedback tends to increase the range in the $\rho$-$T$-phase diagram, where CO-dark gas is found.}
 \label{fig:temp_dens_diag_FB_MHD}
\end{figure*}

In Fig.~\ref{fig:temp_dens_diag_FB_MHD} we show the H$_2$-mass-weighted $n_\rmn{H,tot}$-$T$-phase diagram of the total, CO-dark and CO-bright gas in the runs MC1-HD-FB and MC3-MHD.

\subsection{CO spectra}

\begin{figure*}
 \includegraphics[width=\textwidth]{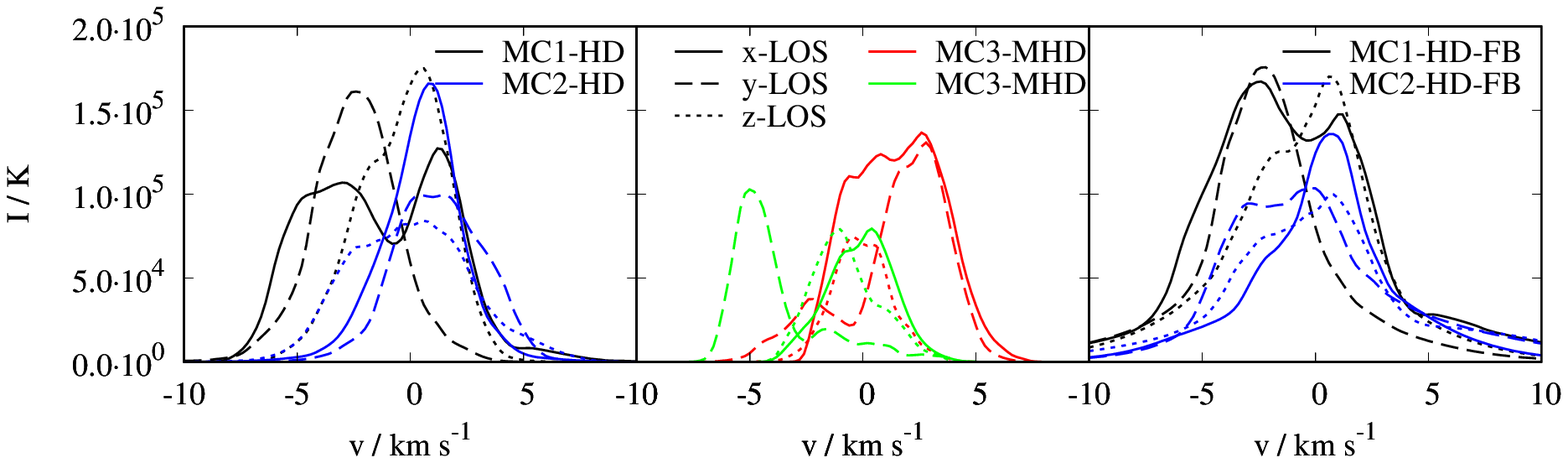}\\
 \caption{CO-spectra of the entire zoom-in region for three different LOS and all six MCs considered at $t_\rmn{evol}$ = 3 Myr. The spectra show a large variety indicating various components along the LOS and optical depth effects.}
 \label{fig:CO-spectra}
\end{figure*}

In Fig.~\ref{fig:CO-spectra} we show the CO spectra of the entire zoom-in regions of the six different simulations for three different LOS at $t_\rmn{evol}$ = 3 Myr. The spectra differ significantly between the different simulations but also for individual cloud when considering different LOS. The spectra show features which can be attributed to multiple velocity components along the LOS but also to optical depth effects towards the line center. As a detailed analysis of the spectra is beyond the scope of the paper, we defer it to a subsequent publication (N\"urnberger et al. in prep.).

%


\bsp	
\label{lastpage}
\end{document}